\documentclass[10pt]{amsart}
\usepackage[english]{babel}
\usepackage{graphicx}

\title[An algebraic approach to discrete time integability]{An algebraic approach to discrete time integrability}
\author[Anastasia Doikou and Iain Findlay]{Anastasia Doikou and Iain Findlay}

\address[A. Doikou] {Department of Mathematics, Heriot-Watt University,
Edinburgh EH14 4AS, and The Maxwell Institute for Mathematical Sciences, Edinburgh, UK}
\email{A.doikou@hw.ac.uk}

\address[I. Findlay] {
}
\email{Iain\_Findlay@yahoo.com}

\usepackage[english]{babel}
\usepackage[utf8]{inputenc}
\usepackage{amsmath, amssymb, bm}
\usepackage{graphicx}
\usepackage{subcaption}
\usepackage[colorinlistoftodos]{todonotes}
\usepackage{indentfirst}

 	\definecolor{coolblack}{rgb}{0.0, 0.18, 0.39}
\definecolor{calpolypomonagreen}{rgb}{0.12, 0.3, 0.17}
\definecolor{cadmiumgreen}{rgb}{0.0, 0.42, 0.24}

\newcommand{\hiddenpower}[2] { \ifnum \numexpr#2=1 #1 \else #1^#2 \fi }
\numberwithin{equation}{section}

\newcommand{\cal}{\mathcal}

\usepackage{xparse}
\ExplSyntaxOn
\newcounter{diff_order}
\newcounter{diff_power}

\newcommand{\rawdiff}[3]
{
	\setcounter{diff_order}{0}
	\clist_map_inline:nn{#3}{\stepcounter{diff_order}}
	
	\frac{\hiddenpower{#1}{\thediff_order} #2}
	{
		\def\old_var{DefaultValue}
		\setcounter{diff_power}{0}
		
		\clist_map_inline:nn{#3}
		{
			\def\new_var{##1}
			\ifnum \thediff_power=0
				\stepcounter{diff_power}
			\else
				\tl_if_eq:NNTF \new_var \old_var
				{\stepcounter{diff_power}}
				{
					#1 \hiddenpower{\old_var}{\thediff_power}
					\setcounter{diff_power}{1}
				}
			\fi

			\def\old_var{##1}
		}
		
		#1 \hiddenpower{\old_var}{\thediff_power}
	}
}
\ExplSyntaxOff

\newlength{\bibitemsep}
\newlength{\bibparskip}
\let\oldthebibliography\thebibliography
\renewcommand\thebibliography[1]{%
  \oldthebibliography{#1}%
  \setlength{\parskip}{\bibitemsep}%
  \setlength{\itemsep}{\bibparskip}%
}

\newcommand{\lb}{\left(}
\newcommand{\rb}{\right)}

\newtheorem{thm}{Theorem}[section]

\newtheorem{rem}[thm]{Remark}

\renewcommand{\sinh}[2][1]{\hiddenpower{\text{sinh}}{#1} \lb #2 \rb}
\renewcommand{\cosh}[2][1]{\hiddenpower{\text{cosh}}{#1} \lb #2 \rb}

\renewcommand{\ln}[1]{\text{ln} \lb #1 \rb}

\begin{document}




%



\begin{abstract}
\noindent We propose the systematic construction
of  classical and quantum two dimensional space-time lattices primarily based on
algebraic considerations, i.e. on the existence of associated $r$-matrices and 
underlying spatial and temporal classical and quantum algebras. This is a  novel
construction that leads to the derivation of fully discrete
integrable systems governed by sets of consistent integrable non-linear space-time difference equations. 
To illustrate the proposed methodology, we derive two versions of 
the fully discrete non-linear Sch\"rodinger type system.
The first one is based on the existence of a rational $r$-matrix, whereas the second one  is the
fully discrete Ablowitz-Ladik model and is associated to a trigonometric $r$-matrix. 
The Darboux-dressing method is also applied for the first discretization  scheme, mostly as a consistency check, 
and solitonic as well as general solutions, in terms of solutions of the fully discrete heat equation, are also derived.
The quantization of the fully discrete systems is then quite natural in this context and 
the two dimensional quantum lattice is thus also examined.
\end{abstract}

\maketitle

\date{}
\vskip 0.4in



\section{Introduction}

\noindent 
The fundamental paradigm  in the frame of classical integrable systems is the AKNS scheme \cite{AKNS1}. 
This offers the main non-relativistic set up, and  is naturally associated to the non-linear Schr\"odinger system (NLS),
the mKdV and KdV equations, and can be also mapped to typical examples of relativistic systems such as the sine-Gordon model.
The AKNS scheme and NLS type hierarchies are among the most widely studied integrable prototypes
(see for instance \cite{AKNS1, AL, Ablo, FordyKulish, Manakov} and \cite{Kulish, KunduRagnisco, Sklyanin}). 
Both continuum and discrete versions
have  been thoroughly investigated from the point of view of the inverse scattering 
method or the Darboux and Zakharov-Shabat (ZS) dressing methods 
\cite{ZakharovShabat1, ZakharovShabat2}, \cite{Ablo}, \cite{Darboux, Degalomba1, Mik1, Rourke, Schiff, Zullo, ADP, DFS1, DoiSkl}, yielding
solutions of hierarchies of integrable non-linear PDEs (ODEs) as well as  hierarchies of  associated Lax pairs. 
Numerous  studies from the Hamiltonian point of view  
in the case of periodic (see for instance \cite{FT, Sklyanincl, STS} and \cite{FredelMaillet, GerIv}) and
generic integrable boundary conditions \cite{Sklyanin2, AvanDoikou1} also exist.
The Hamiltonian or algebraic frame offers the most systematic means for
constructing and studying  classical  integrable systems. 
The potency of the algebraic approach relies  on the existence of a classical 
$r$-matrix that satisfies the classical Yang-Baxter equation. This then signifies the presence
of associated  Poisson structures \cite{Sklyanincl, STS} that naturally lead to sets of 
quantities in involution, i.e. integrals of motion.
Quantization in this context is then quite natural as the classical $r$-matrix is replaced 
by a quantum $R$-matrix that obeys the quantum YBE, and the classical Poisson algebra 
is replaced by a quantum algebra \cite{FadTakRes}. 
The existence of a classical (quantum) $r$-matrix allows also the computation of the time components of the Lax pairs of the hierarchy 
via the fundamental Semenov-Tian-Shansky formula (STS)  \cite{STS}, that involves the $r$ and $L$ matrices. 
This universal formula has been extended  to the case of open boundary conditions 
as well as at the quantum level  \cite{AvanDoikou1, DoikouFindlay1, Korepin}.

In the present investigation we are  proposing the algebraic setting for constructing
``space-time'' discrete integrable systems. The study of fully discrete systems has been a particularly active field in 
recent decades, especially after the prototypical Hirota's works \cite{Hirota} on non-linear partial difference equation, 
leading also to  intriguing connections with quantum integrable systems \cite{Zab, HoKou}, 
(see also \cite{HJN} and references therein).
A fundamental frame for describing  such  integrable systems and the associated 
partial difference equations is  the  so-called consistency approach \cite{NRGO, ABS,  HV}.
These studies have also produced various significant connections with
Yang-Baxter maps and the set theoretic Yang-Baxter equation (also linked to the notion of 
Darboux-B\"acklund transformations) \cite{Pap, Ves}, 
cluster algebras \cite{Cluster, Kedem, HoneKouloukas}, and the concept of algebraic 
entropy (see e.g. \cite{entr1, entr2} and references therein), 
to mention a few. Our approach is mainly based on algebraic considerations and is greatly inspired  
by earlier works on space-time  dualities  \cite{ACDK, AvanCaudrelier, CauKu, DFS1, Findlay} 
and the existence of underlying spatial and temporal Poisson structures.
To illustrate the algebraic approach we present  two  distinct fully discrete versions of the NLS-type hierarchy
based on the existence of classical and quantum $r$-matrices and the
underlying deformed algebras:
1) the fully discrete version of the system  introduced in \cite{KunduRagnisco, Sklyanin} (fully DNLS), 
which is the more natural discretization of 
the NLS-type systems (AKNS scheme generally) from the algebraic point of view, and is associated to a rational $r$-matrix.
2) The fully discrete  Ablowitz-Ladik (AL) model (see e.g. \cite{AL, Ablo,  Kulish}) associated to a trigonometric $r$-matrix. 
Generalized local \cite{Darboux}  transformations are then employed in order to identify solutions 
of the associated fully DNLS  nonlinear partial difference equations as well as to confirm the findings from the algebraic point of view.
When discussing the solutions of the relevant partial difference equations we are primarily
focused on the discrete version of the DNLS hierarchy associated to a rational $r$-matrix (see \cite{Sklyanin} 
and references therein). Note that the DNLS model is a natural integrable 
version of the discrete-self-trapping equation introduced and studied in \cite{DST} 
to model the nonlinear dynamics of small molecules, such as ammonia,
acetylene, benzene, as well as large molecules, such as acetanilide. It is also related to
various physical problems such as arrays of coupled nonlinear
wave-guides in nonlinear optics and quasi-particle motion on a dimer among others. 

We stress that this is the first time to our knowledge that a systematic construction 
of fully discrete space-time integrable 
systems  based on the existence of a classical $r$-matrix
is achieved. This fundamental idea is naturally extended to the quantum case 
and the two dimensional quantum lattice can then be constructed. 
This derivation is based on the existence of copies of two distinct quantum algebras 
associated to spatial and temporal ``quantum spaces" and it is in a manner in the spirit of  constructing 
 higher dimensional quantum lattices via the solution of the tetrahedron equation \cite{Zamo, Bazh} (see also relevant \cite{Bazh2}), 
although in our construction
there is a clear distinction between spatial and temporal quantum algebras.

Let us briefly outline what is achieved in the article:
\begin{itemize}
\item In section 2 we present the spatial and temporal Poisson structures associated to
discrete time integrable systems. In this frame the time components of  Lax pairs, i.e. 
the $V$-operators are required to be representations
of a quadratic Poisson structure, whereas the space components satisfy linear Poisson structures
 in the semi-discrete time case,  and quadratic classical algebras in the  fully discrete case. We first examine
 the semi discrete  time case and we consider the time like approach, i.e. for a given $V$-operator we 
apply the corresponding STS formula \cite{ACDK} and derive
the hierarchy of the  space components of the Lax pairs. This part serves as a predecessor,
providing the main frame to consistently formulate the fully discrete case. After we provide the general algebraic 
set up for fully discrete integrable systems
we examine two prototypical systems that are discretizations of the NLS-type scheme. 
For both examples the time components of the Lax pairs are constructed as representations of the quadratic temporal algebras.
Having identified the Lax pairs we also derive the associated partial difference equations 
 via the fully discrete zero curvature condition.

\item In section 3  the Darboux-B\"acklund methodology is implemented for the fully DNLS system.
The purpose of this section is two-fold : 1)  We extract the Lax pairs for the space-time discretization of NLS 
confirming the findings of the algebraic approach. 2) We derive solutions via certain local Darboux transforms.
More specifically, 
by employing the fundamental Darboux transformation we perform the dressing process and we identify 
the Lax pairs of the discrete hierarchy. Explicit expressions for the first few members are presented and 
the findings of the algebraic approach are confirmed.
Via the fundamental Darboux transform we  also derive two types of discrete  solitonic 
solutions that are the fully discrete analogues of the 
solutions found in  \cite{DoiSkl}. More importantly, with the use of a Toda type 
Darboux matrix we identify generic new solutions (i.e. not only solitonic) of the non-linear partial difference
equations in terms of solutions of the associated linear equations, i.e. the fully discrete heat equation, 
generalizing the findings of \cite{DoiSkl} to the fully discrete case.

\item In section 4 we present the two dimensional quantum lattice. 
In order to be able to build the two dimensional quantum lattice along 
the space and time directions, in analogy 
to the classical case as described in section 2, we introduce the notion of
spatial and temporal  ``quantum spaces''.  Despite the slight abuse of language, 
we employ the notion of quantum spaces
to describe copies of the underlying quantum algebras
when constructing  the corresponding spin chain like systems with $N$ ($M$) sites, 
or quantum spaces, along the space (time) direction. 
This construction is in exact analogy to the classical description.  
The quantum discrete NLS model, associated to the Yangian $R$-matrix,  as well as the 
quantum Ablowitz-Ladik model (or $q$ bosons), associated to a trigonometric $R$-matrix, 
are considered  as our prototypical quantum systems.


\end{itemize}

\section{Discrete time Integrability: algebraic formulation}

\noindent In this section we suggest the algebraic formulation for the construction of discrete time integrable systems.
Specifically, we present the  space and time like Poisson structures associated to
discrete time integrable systems.  We first consider
the semi-discrete  time case, which basically serves as a predecessor of the fully discrete frame.
The Lax pairs are perceived in this context as representations of the  underlying space-time Poisson structures.
The discrete versions of the zero curvature condition provide compatibility conditions among the various fields involved and yield
the associated difference/differential equations.
To explicitly illustrate the proposed methodology
we examine two prototypical systems that are discretizations of the NLS-type scheme and are associated to two distinct 
classical $r$-matrices (rational versus trigonometric). 
For both examples the Lax pairs are constructed as 
representations of the spatial and temporal algebras. Having identified the Lax pairs we derive
the associated partial difference equations from the fully discrete zero curvature condition.

\subsection{The semi-discrete time setting}

\noindent  We first examine the case of  discrete time and continuous space classical integrable systems and
we mainly focus on the time-like algebraic picture. 
The main reason we consider this case first is the fact that the STS formula  is available,
and the hierarchy of associated $U$-operators can be thus systematically derived \cite{ACDK}. 
This will be achieved in the following subsection for a particular example, the semi discrete time NLS system. Such a construction 
will provide a  first indication on the consistent forms of Lax pairs in the fully discrete scenario. Note that in this case we only consider
integrable systems associated to rational $r$-matrices, i.e. the Yangian \cite{Yang}.

Let us first recall the space-like description and we then move on to the time-like picture as discussed in \cite{ACDK, AvanCaudrelier, DFS1}. 
The starting point is the existence of a Lax pair $\Big (U,\ V\Big) $ consisting 
of generic $c$-number 
$d \times d$  matrices (see e.g. \cite{FT}). The Lax pair matrices depend in general on some fields and a spectral parameter, 
and form the auxiliary linear problem:
\begin{eqnarray}
&&\partial_x \Psi(x,a,\lambda) = U(x, a,\lambda) \Psi_a(x,a,\lambda),  \nonumber\\
&& \Psi( a+1,x, \lambda)= V(x, a, \lambda) \Psi(x, a,\lambda), \label{ALP}
\end{eqnarray}
where $ a$ is the discrete time index. Compatibility of the two equations above leads to the discrete 
time zero curvature condition:
\begin{equation}
\partial_x V (x, a,\lambda) = U(x ,a+1, \lambda)V(x,a, \lambda,x) - V(x, a,\lambda, x) U(x,a,\lambda). \label{xzero}
\end{equation}

Before we move on to the algebraic formulation of discrete time integrability let us first introduce some 
useful objects. Let us define the space like monodromy matrix, which is a solution of the first of 
the equations of the auxiliary linear problem (\ref{ALP})
\begin{equation}
T_S(x,y, a, \lambda) =  \overset{\curvearrowleft} {\mathrm P} \mbox{exp}\Big \{ \int_{y}^x\ U_a(\xi, \lambda) d\xi \Big \}, 
~~~~~x>y, \label{pathmono}
\end{equation}
where $ \overset{\curvearrowleft}{\mathrm P}$ denotes path ordered integration. We also define the periodic space transfer matrix as 
${\mathfrak t}_S(a,\lambda) = trT_S(A,-A,a,\lambda)$, then using the discrete time zero curvature condition as well as assuming
periodic space
boundary conditions, or vanishing conditions at $\pm A$, we conclude that ${\mathfrak t}_S(a,\lambda)$ is constant in the discrete time, i.e. 
${\mathfrak t}_S(a,\lambda)={\mathfrak t}_S(a+1,\lambda)$.
A more detailed discussion on the latter statement 
 is provided in the next subsection, where the fully  discrete case is examined.

The $M$-site time monodromy $T_T$, which is a solution to the time part of \eqref{ALP}, is defined as
\begin{equation}
T_T( x, b, a, \lambda) =V(x, b, \lambda) \cdots V(x,a+1,\lambda)\  V(x,a,\lambda ), ~~~~~b>a,\label{tmono}
\end{equation} 
and the time-like transfer matrix is given by ${\mathfrak t}_T(\lambda) = tr T_T(x,M,1,\lambda)$. 
By means of the zero curvature condition (\ref{xzero}) and 
assuming periodic time-like boundary conditions we conclude that ${d{\mathfrak t}_T(\lambda)\over dx} =0$


From the algebraic point of view the fundamental statement is that
the $U$-operator satisfies the linear Poisson structure \cite{FT, Sklyanin}
\begin{equation}
\Big\lbrace U(x,a, \lambda)\ \underset{^{,}}\otimes\   U(y, a,\mu) \Big\rbrace_{S} = \Big[ r(\lambda - \mu),\  
U(x,a, \lambda) \otimes I + I \otimes U(y,a, \mu) \Big] \delta(x - y), \label{eq:LinAlg}
\end{equation}
where $I$ is in general the $d \times d$ identity matrix,  the subscript $_S$ 
denotes space-like Poisson structure, and
the $r$-matrix is a solution of the classical Yang-Baxter equation \cite{STS}, 
\begin{equation}
\big [r_{12}(\lambda_1-\lambda_2),\ r_{13}(\lambda_1)\big ] + \big [r_{12}(\lambda_1-\lambda_2),\
 r_{23}(\lambda_2)\big ]+\big [r_{13}(\lambda_1),\ r_{23}(\lambda_2)\big ] = 0. \label{cybe}
\end{equation}
The  $r$-matrix acts on ${\cal V} \otimes {\cal V}$, ${\cal V}$ is a $d$ dimensional vector space in general, 
and in the index notation $r_{12} = \sum_{ij} r(ij|kl) e_{ij} \otimes e_{kl} \otimes I$, similarly for $r_{23}$ and $r_{13}$, and
$e_{ij}$ are in general $d \times d$ matrices with elements $(e_{ij})_{kl} = \delta_{ik} \delta_{jl}$. 

In \cite{ACDK}, where the continuum space-time scenario was examined, 
it was assumed that $V,$ as well as $U$ satisfy linear
Poisson structures  (see also \cite{AvanCaudrelier} on further emphasis on the algebraic/$r$-matrix description).
Indeed, it was noticed  that the time-like Poisson bracket could be constructed 
from a corresponding  algebraic expression regarding the time component of the Lax pair. 

Here we assume time-like discretization and introduce time-like indices $a,\  b$, then the $V$-operator in (\ref{ALP}) is required 
to satisfy the quadratic algebra:
\begin{equation}
	\Big\lbrace V(x,a,\lambda)\ \underset{^{, }}\otimes\  V(x,b,\mu) \Big\rbrace_T= \Big[ r(\lambda - \mu),\ 
V(x,a,\lambda) \otimes V(x,b,\mu) \Big] \delta_{ab}, \label{Dual2}
\end{equation}
where $r$ is the same classical $r$-matrix as in (\ref{eq:LinAlg}), and the subscript $_T$ denotes 
the time-like Poisson structure. Both space and time Sklyanins bracket's (\ref{eq:LinAlg}) and (\ref{Dual2}) 
are the typical Poisson structures on
the $LGL_d$ loop group. 

The time like mondromy matrix (\ref{tmono}) satisfies the quadratic algebra (we write for simplicity 
$T_T(x,M,1,\lambda) = T_T(\lambda)$)
\begin{equation}
\Big \{ T_{T}(\lambda)\  \underset{^{,}}\otimes\ T_{T}(\mu) \Big \}_T= \Big [ r(\lambda -\mu),\ 
T_{T}(\lambda)\otimes  T_{T}(\mu) \Big ]. \label{quad}
\end{equation}
Consequently, one obtains commuting operators, with respect to the time-like Poisson structure 
$\Big \{trT_T(\lambda),\ trT_T(\mu) \Big \}_T=0.$

\subsubsection{Deriving $V$-operators}
\noindent  Our main  objective now is to identify the form of the time components of the Lax pairs,
 i.e. the $V$-operators for algebras associated to the rational 
$r$-matrix \cite{Yang},
\begin{equation}
r(\lambda) ={1\over\lambda}\ \sum_{i, j=1}^{d} e_{ij} \otimes e_{ji}. \label{YangS}
\end{equation}
The quantity $\sum_{i, j} 
e_{ij} \otimes e_{ji}$  is the so called permutation operator.
We express the $V$-operator in the following generic form 
as a finite $\lambda$ series expansion
\begin{equation}
V^{(k)}(\lambda)=  \sum_{m=0}^{k} \lambda^{m}{\mathcal Y}^{(m, k)}. \label{str}
\end{equation}
In the case we examine here, i.e. the DNLS hierarchy we consider  
${\cal Y}^{(k,k)}={\cal D}= \mbox{diag}(1,\ 0,\ldots, 0)$  ($d-1$ zero diagonal zero entries in general). 
${\cal Y}^{(m,k)}$ are in general $d\times d$ matrices to be identified algebraically.
Note that for ${\cal Y}^{(k)} = I$ ($I$ is the $d\times d$ identity matrix) we essentially deal 
with the classical version of the $\mathfrak{gl}_{d}$ Yangian (here we focus on $d=2$). 
More generally in the Yangian case ${\cal Y}^{(k,k)}$ can be a constant
non-singular matrix.

We impose the following two fundamental assumptions  in order to identify each $V^{(k)}$ of the generic form (\ref{str}).
\\

\noindent $\bullet$ {\bf The basic assumptions}
\begin{enumerate}
\item Each $V^{(k)}$ of the form (\ref{str}) satisfies the quadratic algebra (\ref{Dual2}).
\item $\mbox{det} V^{(k)} = \lambda^k +\sum_{n=0}^{k-1}a_n \lambda^n$.
\end{enumerate}
$ $

\noindent From assumption 1 and the general form of the $V^{(k)}$-operator (\ref{str}), the following Poisson relations emerge, 
being the classical analogues of the Yangian $\mathfrak{gl}_d$ (${\cal Y}^{(k,k)}$ is a constant (non-dynamical)  matrix)
\begin{eqnarray}
&& \Big \{ {\cal Y}^{(m-1,k)} \underset{^{, }}\otimes {\cal Y}^{(0,k)}\Big \}_T = \Big [ {\cal P},\ {\cal Y}^{(m,k)} \otimes {\cal Y}^{(0,k)} \Big ] \nonumber \\
&&  \Big \{ {\cal Y}^{(m-1,k)} \underset{^{, }}\otimes {\cal Y}^{(l,k)}\Big \}_T-  \Big \{ {\cal Y}^{(m,k)} \otimes {\cal Y}^{(l-1,k)}\Big \}_T=
 \Big [ {\cal P},\ {\cal Y}^{(m,k)} \otimes {\cal Y}^{(l,k)} \Big ], 
\end{eqnarray}
where  $m,l \in \{1, \ldots, k \}$ and  ${\cal P}= \sum_{i, j=1}^d  e_{ij} \otimes e_{ji}$ is the permutation operator.

\noindent In the language of dressing Darboux transform assumption 2
is equivalent to saying that the determinant of $V^{(k)}$ is independent of the fields, i.e. 
the determinant of the ``dressed'' $V$-operators should be equal to the one of the 
``bare'' operators (free of fields). By employing the two fundamental assumptions above
we can then express all ${\cal Y}^{(m,k)}$ in terms of some ``fundamental'' fields (see also \cite{DFS1} for a relevant discussion), 
that satisfy certain basic Poisson relations. The problem then reduces into classifying representations of the classical algebra (\ref{Dual2}) 
of the general structure (\ref{str}).

Given the operator $V^{(k)}$ we can then apply the time-like  STS formula \cite{STS, ACDK}, 
which is valid for continuum $x$ \cite{DFS1}, derive the hierarchy of associated 
$U$-operators and extract in turn the hierarchy of non linear integrable ODEs. 
This will be achieved in the next subsection. 
Notice that extra compatibility conditions emerge from the discrete time zero curvature 
condition ensuring the consistency of our construction.
Let us now focus on the first two members of the hierarchy, and identify $V^{(1)},\ V^{(2)}$ from the algebraic point of view. 
In the continuum time situation for each time flow $t_k$ a corresponding $V^{(k)}(x, t_k, \lambda)$ exists. 
In analogy, in the discrete case our notation will be $V^{(k)}(x, a_k, \lambda)$ for each discrete time 
index $a_k$. We shall drop the sub-index $k$ henceforth for brevity.\\

\noindent {\bf The $V^{(1)}$-operator.} The first non-trivial $V$-operator is linear in $\lambda$ and is associated to a discrete time version 
of the transport equation,
\begin{eqnarray}
 V^{(1)}(x,a,\lambda) &=& \lambda {\cal D} + {\cal Y}^{(0,1)}\nonumber\\ &=&  \lb \begin{matrix}
		 \lambda +  {\mathbb N}^{(1)}_{a}(x) & \hat u_{a}(x)  \\
		  u_{a}(x)  &  1
	\end{matrix} \rb,  \label{NLSb1}
\end{eqnarray}
where from the condition $\mbox{det} V^{(1)} = \lambda+ 1$ we obtain
\begin{equation}
{\mathbb N}_a^{(1)}(x) = 1 + \hat u_a(x) u_a(x). \label{nn}
\end{equation} 
Note that the $x$ dependence in the fields is always implied even if omitted for brevity.
Due to the fact that $V^{(1)}$ satisfies the quadratic algebra (\ref{Dual2}) 
we derive the Poisson relations:
\begin{equation}
\Big \{ {\cal Y}^{(0,1)}\  \underset{^{, }}\otimes {\cal Y}^{(0,1)} \Big \}_T = 
{\cal P} \Big ({\cal D} \otimes {\cal Y}^{(0,1)} -  {\cal Y}^{(0,1)} \otimes {\cal D}\Big ),
\label{basica1}
\end{equation}
where ${\cal P} = \sum e_{ij} \otimes e_{ji}$ is the permutation operator. 
Hence, the Poisson relations for the fields  follow
\begin{equation}
\Big \{ u_a(x),\ \hat u_b(x) \Big \}_T = \delta_{ab}, ~~~ \Big \{\hat u_a(x),\ {\mathbb N}_b^{(1)} \Big \} _T=\hat u_a(x)\delta_{ab},
~~~\Big \{ u_a(x),\ {\mathbb N}_b^{(1)} \Big \}_T = u_a(x)\delta_{ab}. \label{cla1}
\end{equation}
The field ${\mathbb N}^{(1)}$ (\ref{nn}) is apparently compatible with the classical  algebra (\ref{cla1}). $\square$
 \\

\noindent {\bf The $V^{(2)}$-operator.} We now derive the $V$-operator quadratic in $\lambda$:
\begin{eqnarray}
 V^{(2)}(x, a, \lambda) &=& \lambda^2{\cal D} + \lambda {\cal Y}^{(1,2)}(x,a) + {\cal Y}^{(0,2)}(x,a)  \nonumber\\  
&=& \lb \begin{matrix}
		\lambda^2+ \lambda {\mathbb N}^{(2)}_a + {\mathbb A}_a & \lambda \hat u_a+ {\mathbb  B}_a \\
		\lambda  u_a + {\mathbb  C}_a & {\mathbb  D}_a
	\end{matrix} \rb. \label{NLSb2}
\end{eqnarray}
Requiring that $\mbox{det} V^{(2)} = \lambda^2+ 1$ we conclude that the algebraic quantities ${\mathbb N},  {\mathbb A}, {\mathbb D}$ 
are expressed in terms of the fundamental fields $u,\ \hat u,\ {\mathbb B},\  {\mathbb C}$ as
\begin{eqnarray}
{\mathbb N}^{(2)}_a = {u_a {\mathbb B}_a  +\hat u_a{\mathbb C}_a \over 1 +\hat u_a u_a}, ~~~~~
{\mathbb  A}_a = {1 + {\mathbb B}_a {\mathbb C}_a \over 1 + \hat u _a u_a}, ~~~~~~{\mathbb  D}_a = 1 +\hat u_a u_a. \label{FieldsA0}
\end{eqnarray}
Requiring also that $V^{(2)}$ satisfies the time-like Poisson structure  
(\ref{Dual2}) we then produce the Poisson relations
\begin{eqnarray}
&&\Big \{ {\cal Y}^{(1,2)}\  \underset{^{, }}\otimes {\cal Y}^{(1,2)} \Big \}_T= {\cal P} \Big ( {\cal D} \otimes {\cal Y}^{(1,2)} -  {\cal Y}^{(1,2)} \otimes {\cal D}\Big ),\label{basicb1}\\
&&
\Big \{ {\cal Y}^{(1,2)}\  \underset{^{, }}\otimes {\cal Y}^{(0,2)} \Big \}_T= {\cal P} \Big ( {\cal D} \otimes {\cal Y}^{(0,2)} -  {\cal Y}^{(0,2)} \otimes {\cal D}\Big ),\label{basicb2}\\
&&
\Big \{ {\cal Y}^{(0,2)}\  \underset{^{, }}\otimes {\cal Y}^{(0,2)} \Big \}_T = {\cal P} \Big ( {\cal Y}^{(1,2)} \otimes {\cal Y}^{(0,2)} -  
{\cal Y}^{(0,2)} \otimes {\cal Y}^{(1,2)} \Big ). \label{basicb3}
\end{eqnarray}
and hence the time-like algebra for the fields (we only write below the non zero commutators for the fundamental fields, 
see also \cite{ACDK}, see also Appendix B for the corresponding quantum algebra relations):
\begin{eqnarray}
&&  \Big \lbrace u_a(x),\ {\mathbb B}_b(x) \Big \rbrace_T  = -\Big \lbrace \hat  u_a(x),\ {\mathbb C}_b(x) \Big \rbrace_T  = 
\Big (1 + \hat u_a(x) u_a(x)\Big )\delta_{ab}, \nonumber\\
&&  \Big \{{\mathbb B}_a(x),\ {\mathbb C}_b(x) \Big\}_T = -\Big( u_a(x) {\mathbb B}_a(x) +
\hat u_a(x){\mathbb  C}_a(x) \Big) \delta_{ab}. \label{Dual2alg}
\end{eqnarray}
The quantities defined in (\ref{FieldsA0}) are compatible with the algebra  (\ref{Dual2alg}). $\square$

It is worth noting that in the space-like formulation the $U$-matrix (\ref{ALP}) is the starting point and the conserved 
quantities as well as the hierarchy of $V$-operators emerge from it \cite{STS,  DoFioRa}. In the time-like approach 
on the other hand the starting point is some $V$-operator, and from this the time-like conserved quantities as well as the 
$U$-hierarchy are derived \cite{ACDK}.
In the next subsection,  we focus only on time-like Poisson structures thus we drop the subscript $_T$ whenever this applies.

\subsubsection*{The discrete time-like Lax pair hierarchy}
\label{sec:Periodic}

\noindent  
Here we exclusively discuss the time-like case, and extract the associated charges in 
involution as well as the hierarchy of $U$-operators.
The generating function of the hierarchy of the local conserved 
quantities\footnote{``Conserved'' with respect with respect to spatial variations for the
 monodromy matrix built using $ V $.}    
associated to the system is given by:
\begin{equation}
	\mathcal{G}(\lambda) = \ln{tr \big (T (\lambda)\big )}, \nonumber
\end{equation}
where $T(\lambda) = T_T( M,1,\lambda)$ the time-like monodromy (\ref{tmono}) ($x$ dependence is implied).

We may also  derive the generating function that provides the hierarchy of $U$-operators associated to 
each one of the time-like Hamiltonians.  Indeed, taking into consideration the zero curvature condition as 
well the time-like Poisson structure satisfied  by $V$ one can show 
that the generating function of the 
$U$-components of the Lax pairs is given by the time-like analogue of the STS formula  (see also \cite{ACDK} for a more detailed derivation)
\begin{align}
{\mathbb U}_2(a, \lambda, \mu) &= \mathfrak{t}^{-1}(\lambda) tr_1\Big (T_1(M, a, \lambda) r_{12}(\lambda - \mu) 
T_1(a-1, 1, \lambda)\Big ), \label{eq:UGen}
\end{align}
where recall the time-like monodromy matrix defined in (\ref{tmono}) for $b> a$. 
We also  introduce the index notation: 
$A_1 = A \otimes I$ and $A_2 = I \otimes A$ for any $d \times d$ matrix $A$, $I$ is the 
$d \times d$ identity matrix, and $r$
acts on ${\cal V} \otimes {\cal V}$ (${\cal V}$ is the $d$ dimensional vector space).
In the case where the $r$-matrix is the 
Yangian (\ref{YangS}) the latter expression (\ref{eq:UGen}) reduces to
\begin{eqnarray}
{\mathbb U}(a, \lambda, \mu) &=& {\mathfrak{t}^{-1}(\lambda) \over \lambda -\mu}\  T(a-1, 1, \lambda) \ T(M, a, \lambda) .
\label{eq:UGen2}
\end{eqnarray}

We restrict our attention now on the hierarchy  associated to $V^{(2)}$ (\ref{NLSb2}). Indeed, 
expanding the monodromy matrix (\ref{tmono}) constructed by the $V$-operator (\ref{NLSb2}), 
in powers of ${1\over \lambda}$, we obtain the associated charges in involution. We report below 
the first couple of conserved  quantities:
\begin{equation}
\begin{aligned}
	H^{(1)} &= \sum_{a=1}^M {u_a {\mathbb B}_a + \hat u_a {\mathbb C}_a \over 1 + u_a \hat u_a} , \\
	H^{(2)} &=  \sum_{a=1}^M  \Big ( \hat u_a u_{a-1}+ {1 + {\mathbb B}_a {\mathbb C}_a \over 1 + u_a \hat u_a} -
{1\over 2} \big( {u_a {\mathbb B }_a + \hat u_a {\mathbb C}_a \over 1 + u_a\hat u_a}\big)^2 \Big ) \\  \ldots
\end{aligned} \label{Hams}
\end{equation}
In fact, $H^{(2)}$ is the Hamiltonian of the  semi discrete time  NLS system.

In addition to the derivation of the time-like charges in involution above we can also compute the corresponding 
$U$-operators of the time-like hierarchy via the expansion in  powers 
of ${1\over \lambda}$  of (\ref{eq:UGen2}).  The pair $\Big (U^{(k)}, V^{(2)} \Big ) $ gives rise to the same equations of motion 
as Hamilton's equations with the Hamiltonian $ H^{(k)} $ associated to the $x_k$ flow.

We provide below the first few members of the series expansion of ${\mathbb U}$ corresponding to the charges (\ref{Hams})
\begin{equation}
	U^{(1)}(\lambda) = \lb \begin{matrix}
		1 & 0 \\
		0 & 0
	\end{matrix} \rb, ~~~~
	U^{(2)}(x, a,\lambda) = \lb \begin{matrix}
		\lambda & \hat u_a(x)\\
		u_{a-1}(x)& 0
	\end{matrix} \rb, 
~~\ldots 
\label{Uops}
\end{equation}
We focus on the second member  of the hierarchy, which is going to give an integrable  time discertization of the NLS model.
Note that the $U^{(2)}$-operator of the system under study, satisfies the algebra (\ref{eq:LinAlg}), thus the space-like Poisson structure 
for the fields is given by:
\begin{equation}
\Big \lbrace u_{a-1}(x),\ \hat u_{a}(y)\Big \rbrace_S=\delta(x-y).
\end{equation}

Having identified both the charges in involution as well as the various $U$-operators, we focus on the 
second member of the hierarchy. In particular, let us obtain
via the Hamiltonian $H^{(2)}$ (and the time-like Poisson relations)  and/or the Lax pair $\Big ( U^{(2)},\ V^{(2)}\Big )$ 
the corresponding equations of motion.  
Equations (\ref{f1}), (\ref{f3}), via the definition of ${\mathbb N}^{(2)}$ (\ref{FieldsA0}), lead to
\begin{equation}
{\mathbb B}_a = {\partial_x\hat u_a - \hat u_a^2 \partial_x u_a \over 1-u_a \hat u_a}, ~~~~~
{\mathbb C}_a = { u_a^2 \partial_x \hat u_a - \partial_x u_a \over 1-u_a \hat u_a}.
\label{FieldsBp}
\end{equation}
Also, from the zero curvature condition we obtain the following constraints
\begin{equation}
\partial_x {\mathbb B}_a = \hat u_{a+1}{\mathbb D}_a - {\mathbb A}_a \hat u_a, ~~~~\partial_x {\mathbb C}_a  = u_a {\mathbb A}_a 
- {\mathbb D}_a u_{a-1}. \label{FieldsCp}
\end{equation}
Given that ${\mathbb A},\ {\mathbb D},\ {\mathbb N}^{(2)}$ (\ref{FieldsA0}), and ${\mathbb B},\ {\mathbb C}$ (\ref{FieldsBp})
are expressed in terms of the fundamental fields $u_a,\ \hat u_a$ and their $x$-derivatives, equations (\ref{FieldsCp})
are the equations of motion for the fundamental fields
fields $u_a,\ \hat u_a$.  

As shown above the Lax pair $\Big (U^{(2)},\ V^{(2)}\Big )$  produces the discrete time 
analogue of the NLS equation, whereas
the Lax pair  $\Big (U^{(2)},\ V^{(3)}\Big )$ is expected to yield the discrete time complex mKdV 
equations  in analogy to the findings of \cite{DFS1, DoiSkl} (see also \cite{CJM} on the mKdV Lax pair). 
The algebraic derivation of $V^{(3)}$ is not included in our computations here as it is quite involved 
and will be presented elsewhere.

\subsection{The fully discrete setting}

\noindent We come now to our primary objective, which is the derivation and study of fully discrete integrable systems. 
In this frame the notion of space-time duality will be more natural given that space and time are at equal footing,
in exact analogy to the continuous space-time picture \cite{ACDK, DFS1}. 
We are going to describe the problem algebraically, whereas
 in the subsequent subsection we apply the fully discrete dressing 
process as a further consistency check on the derivation of  the associated Lax pairs. 

Consider the fully discrete  Lax pair $\Big (L,\ V \Big )$ that  depends on the fields and some spectral parameter.
Let also $n$ denote a discrete 
space index, and $a$  a discrete time one, then
the fully discrete auxiliary linear problem takes  the form:
\begin{eqnarray}
&&\Psi(n+1,a,\lambda) =L(n,a,\lambda) \Psi(n,a,\lambda) \\ \label{aux1}
&& \Psi(n,a+1,\lambda)= V(n,a,\lambda) \Psi(n,a,\lambda). \label{aux2}
\end{eqnarray}
Consistency of the two equation of the auxiliary linear problem lead to the fully discrete equations of motion 
(the fully discrete analogue of the zero curvature condition):
\begin{equation}
V(n+1,a,\lambda) L(n,a,\lambda) = L(n,a+1,\lambda) V(n,a,\lambda). \label{full_zero}
\end{equation}
In this context  both discrete space and time are at equal footing as in the continuous case \cite{ACDK}.

We consider the space-like monodromy matrix defined at some discrete time $a$ as
\begin{equation}
T_S(n,m, a, \lambda) = L(n,a,\lambda) \cdots L(m+1, a,\lambda )L(m,a,\lambda), ~~~n>m, \label{space1}
\end{equation}
and the space-like transfer matrix is defined as ${\mathfrak t}_S(a,\lambda) = tr T_{S}(N,1, a, \lambda)$.
Recall  that in general $L,\ V$ are $d \times d$ matrices and the trace above is defined
with respect to the $d$ dimensional (auxiliary) space.
By virtue of the fully discrete zero curvature condition we show that ${\mathfrak t}_{S}(a, \lambda) = {\mathfrak t}_{S}(a+1, \lambda)$,  i.e.
the transfer matrix  is a constant with respect to the discrete time. Indeed, consider $t_S(a+1, \lambda)$, also 
from (\ref{full_zero}) we have that $L(n,a+1)= V(n+1,a) L(n,a) V^{-1}(n,a)$, then
\begin{equation}
{\mathfrak t}_{S}(a+1, \lambda) =  tr\Big (V(N+1,a,\lambda)  L(N,a,\lambda) \cdots 
L(1,a,\lambda) V^{-1}(1,a,\lambda) \Big ). \label{const1}
\end{equation}
Assuming periodic space like boundary conditions, i.e. $V(N+1,a,\lambda) =V(1,a, \lambda)$, and recalling (\ref{const1}) 
and the definition of the space-like transfer matrix we conclude that ${\mathfrak t}_{S}(a+1, \lambda)= {\mathfrak t}_{S}(a, \lambda)$.
The $\lambda$-series expansion of the transfer provides naturally
the conserved quantities of the system with the respect to the discrete time denoted by the index $a$. Note that the continuous limit of  (\ref{space1}),
($L(n, \lambda) \to I+ \delta U(n, \lambda)$), provides the solution of the space part of the discrete time auxiliary linear problem 
of the previous section given by (\ref{pathmono}).
 
Similarly, let us consider the time-like transfer matrix defined for any space index $n$ as ${\mathfrak t}_T(n,\lambda) = trT_T(n, M,1,\lambda)$, 
where the time monodromy matrix $T_T$ in given by (\ref{tmono}). Through (\ref{full_zero}) $V(n+1,a,\lambda)= 
L(n, a+1,\lambda) V(n,a,\lambda) L^{-1}(n,a,\lambda) $ 
and assuming time like boundary conditions $L(n, M+1,\lambda) =L(n,1,\lambda)$ we conclude that  the transfer matrix
is invariant with respect to the discrete space  indexed by $n$, i.e. ${\mathfrak t}_T(n+1, \lambda)={\mathfrak t}_T(n, \lambda)$,
i.e suitable expansion in powers of $\lambda$ produces the hierarchy of associated invariants for the system with respect
to the discrete space characterized by the index $n$.
In the continuous space limit the latter reduces to ${d{\mathfrak t}_{T}(x, \lambda)\over d x} =0$
(see also comments at the beginning of section 2).


We graphically represent the Lax pair $\Big (L,\ V\Big )$ in our fully discrete set up as:

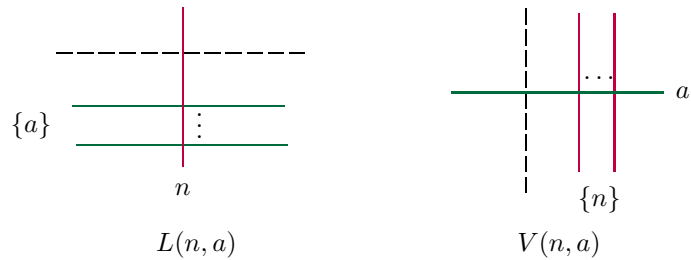
\vspace{1.2cm}
\begin{center}
\begin{picture}(1,1)
	\multiput(-80,10)(8,0){12}{\line(-1,0){6}}
{\color{cadmiumgreen}\put(-80,-10){\line(1,0){80}}}
{\color{cadmiumgreen}\put(-82,-25){\line(1,0){80}}}
{\color{purple}\put(-45,27){\line(0,-2){60}}}
\put(-40,-22){$\vdots$}
	\put(-48,-44){$n$}
\put(-110,-20){$\{a\}$}
\put(-55, -65){$L(n,a)$}
\multiput(85,-43)(0,8){9}{\line(0,1){6}}
{\color{purple}\put(105,25){\line(0,-2){60}}}
{\color{purple}\put(115,25){\line(0,-2){60}}}
{\color{cadmiumgreen}\put(50,-5){\line(1,0){80}}}
\put(100,0){$\ldots$}
\put(98,-48){$\{n\}$}
\put(135,-7){$a$}
\put(75, -65){$V(n,a)$}
\end{picture}
\end{center}

\vspace{2.7cm}

\noindent The set of time indices $\{a\}\equiv \{a,\ a-1, \ldots,  a-k+1 \}$  denotes  discrete time dependence in $L$, 
which is usually implicit. The integer $k$
depends on the form of the $L$ operator. 
Similarly, the set of  space like indices  $\{n\}\equiv \{n,\ n-1, \ldots,  n-l+1\}$ denotes discrete space dependence in $V$, 
which is usually implicit.
In the cases considered here $k=l=2$. The dashed line represents the $d$ dimensional ``auxiliary'' space 
of the Lax pairs (in the examples that follow $d=2$.) 

Next we graphically represent the space-like monodromy:

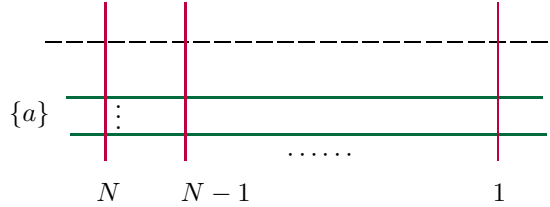
\vspace{1.0cm}

\begin{center}
\begin{picture}(1,1)
	\multiput(-82,15)(8,0){24}{\line(-1,0){6}}
{\color{cadmiumgreen}\put(-80, -6){\line(1,0){180}}}
{\color{cadmiumgreen}\put(-82,-20){\line(1,0){180}}}
	{\color{purple}\put(-72,30){\line(0,-1){60}}}
{\color{purple}\put(-45,30){\line(0,-2){60}}}
{\color{purple}	\put(70,30){\line(0,-1){60}}}
\put(-75,-18){$\vdots$}
	\put(-10,-30){$\cdots\cdots$}
	\put(-82,-45){$N$}
	\put(-50,-45){$N-1$}
	\put(68,-45){$1$}
\put(-115,-15){$\{a\}$}
\end{picture}
\end{center}
\vspace{1.4cm}

\noindent The space-like monodromy $T_S(a)$, corresponds to a one dimensional 
$N$-cite space-like lattice  at a given discrete time $a$. 
The space transfer matrix is defined after taking the trace over the auxiliary space 
resulting in periodic space boundary conditions,  that is the space transfer matrix  is graphically depicted 
by a cylinder, i.e. consider the first and $N^{th}$ site  in the figure above to coincide.
The time-like monodromy is the vertical analogue of the above figure and represents an one
dimensional time $M$-site time-like lattice, 
for a given space index $n$. We basically consider a  90 degrees rotation of the figure 
above and replace the spatial indices$\{ 1, \ N\}$ with temporal ones $\{ 1, \ M\}$, 
and the fixed index $a$ with $n$ (also the colors are interchanged accordingly: 
green $\leftrightarrow$ purple, 
i.e. horizontal lines green and vertical lines purple).

The auxiliary space does not appear in the two dimensional lattice that is graphically 
depicted below for a given lax pair $\Big(L,\ V \Big )$:

\vskip 2.5cm
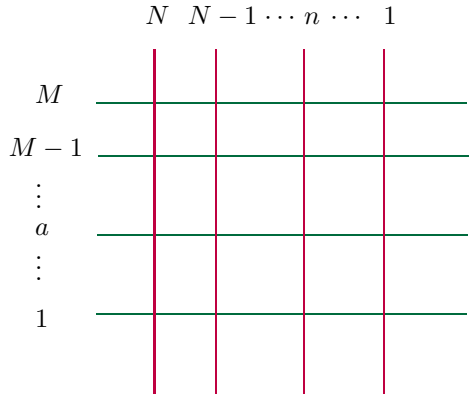
\begin{center}
\begin{picture}(1,1)
{\color{cadmiumgreen} \put(-82,-60){\line(1,0){140}}}
{\color{cadmiumgreen} \put(-85,-30){\line(1,0){140}}}
{\color{cadmiumgreen} \put(-88,0){\line(1,0){140}}}
{\color{cadmiumgreen} \put(-92,20){\line(1,0){140}}}
	\put(-115,-20){$\vdots$}
\put(-115,-47){$\vdots$}
	\put(-115,-65){$1$}
\put(-115, -30){$a$}
	\put(-125,0){$M-1$}
\put(-115,20){$M$}
{\color{purple}\put(-70,-90){\line(0,1){130}}}
{\color{purple}  \put(-50,-90){\line(0,1){130}}}
{\color{purple} \put(10, -90){\line(0,1){130}}}
\put(-35,50){$\cdots$}
\put(-10,50){$\cdots$}
	\put(10,50){$1$}
\put(-20,50){$n$}
	\put(-64,50){$N-1$}
	\put(-80,50){$N$}
{\color{purple} \put(-20,-90){\line(0,1){130}}}
\end{picture}
\end{center}

\vskip 3.70cm


\noindent The figure above should be carefully interpreted, especially  when referring to monodromies and transfer matrices. More specifically,
for a fixed time index $a$ we focus on the space-like monodromy/transfer matrix (\ref{space1}) (horizontally), 
and the respective space-like discrete system whereas,
in the time-like situation the space index $n$ is fixed and we focus on the time-like monodromy/transfer matrix (\ref{tmono})  (vertically) 
(see also relevant  comments on 
the ``conservation''  laws discussed earlier in this section).  The space and time monodromies can be seen as 
horizontal  and vertical ``stripes''  respectively in the two dimensional lattice above.
The latter interpretation applies also in the continuum scenario on the  $x-t$
plane \cite{ACDK, AvanCaudrelier} when considering the  corresponding continuous monodromies. 
The clear distinction between space and time indices  becomes  more transparent below
when presenting the algebraic formulation of the problem.
When considering a given Lax pair and the fully discrete zero curvature condition in order 
to extract the space time difference equations
the two dimensional  lattice is interpreted in the usual sense as the discretization of the $x-t$ plane.

Let us now focus on the algebraic  formulation of fully discrete integrable systems.
The key object  in describing the space-like discrete picture is the $L$ operator, which satisfies the quadratic Poisson structure
\begin{equation}
\Big \{ L(n,a,\lambda) \underset{^{,}}\otimes L(m,a,\mu)\Big \}_S = \Big [r(\lambda- \mu),\ L(n,a,\lambda) 
\otimes L(m,a,\mu) \Big ] \delta_{nm}\label{Poisson2}
\end{equation}
$\lambda,\ \mu$ are spectral parameters, and the $r$-matrix satisfies the classical Yang-Baxter equation. 

Similarly to the semi-discrete time case described in the preceding subsection we require that the time component $V$ of the Lax pair
satisfies the time-like Poisson structure:
\begin{equation}
\Big \{ V(n,a,\lambda) \underset{^{,}}\otimes V(n,b,\mu)\Big \}_T = \Big [r(\lambda- \mu),\ V(n,a,\lambda) 
\otimes V(n,b,\mu) \Big ] \delta_{ab}.\label{Dual2b}
\end{equation}
The classical $r$-matrix is the same as the one of the space-like algebra (\ref{Poisson2}).

The involution of the charges produced by the space and time-like transfer matrices is  guaranteed by the
existence of the Poisson structures (\ref{Poisson2}) and (\ref{Dual2b}). 
Indeed, the monodromies (\ref{space1}) and (\ref{tmono}) satisfy 
(\ref{Poisson2}) and (\ref{Dual2b}) respectively, and  thus 
the corresponding transfer matrices are in involution for different spectral parameters: 
$\big \{{\mathfrak t}_S(\lambda),\  {\mathfrak t}_S(\lambda') \big \}_S =
\big \{ {\mathfrak t}_T(\lambda),\  {\mathfrak t}_T(\lambda')\big \}_T=0$.
This fact stipulates the existence of extra continuous dynamical parameters (underlying continuous ``time'') 
in accordance to Hamilton's equations. The associated hierarchies of the continuous time components of Lax pairs
can then be obtained via the STS formula for both discrete space-like or time-like systems constructed as described above.

We focus now on two distinct versions of the fully discrete NLS 
model associated to rational and trigonometric 
classical $r$-matrices respectively.

\subsubsection{The fully discrete  NLS model}

\noindent We first examine the fully discrete version of the NLS model associated to the Yangian $r$-matrix.
The $L$ operator of the discrete NLS-type hierarchy (\ref{DNLS}) is given as \cite{KunduRagnisco, Sklyanin}
\begin{equation}
 L(n,a,\lambda) = \lb \begin{matrix}
		\lambda+ {\mathbb N}_{na} & {\mathrm X}_{na}\\
		{\mathrm Y}_{na-1} &1
	\end{matrix} \rb, \label{DNLS}
\end{equation}
where ${\mathbb N}_{na} = \theta+ {\mathrm X}_{na}{\mathrm Y}_{na-1} $, $\theta$ is an arbitrary constant.
The Lax operator  satisfies Sklyanin's bracket (\ref{Poisson2}). Notice that the discrete time dependence in 
$L$ is fully justified by the time-like derivation of $U^{(2)}$ via the STS formula in the previous subsection ($U^{(2)}$ 
is the continuum space  limit of $L$). 
The Poisson structure (\ref{Poisson2}) leads to the following Poisson relations among the fields
\begin{eqnarray}
&& \Big\{ {\mathrm X}_{na},\ {\mathrm Y}_{ma-1} \Big \}_S =-\delta_{nm}, ~~~~
\Big \{{\mathrm X}_{na}, \ {\mathrm X}_{ma} \Big\}_S= \Big \{{\mathrm Y}_{na}, \ {\mathrm Y}_{ma} \Big\}_S =0. \label{dn1}
\end{eqnarray}

In analogy to the semi-discrete time case described in the previous section we consider the following generic form for the $V$-hierarchy
\begin{equation}
V^{(k)}(n,a,\lambda)= \lambda^k{\cal D} + \sum_{l=0}^{k-1} \lambda^{l} {\cal Y}^{(l,k)}(n,a), \label{str2}
\end{equation}
where ${\cal D}= \mbox{diag}(1,\ 0)$. To be precise in our notation we should write:\\ $V^{(k)}(n, a_1, a_2,...,a_k..)$, 
however for simplicity we suppress the time-like indices $a_l,\ l \neq k$ and we instead write $V^{(k)}(n,a)$. 
We also require that $\mbox{det} V^{(k)} = \lambda^k +1$, and
 all $V^{(k)}(n,a)$ satisfy the quadratic algebra  (\ref{Dual2b}),
with the same $r$-matrix as in (\ref{Poisson2}).
Then  all ${\cal Y}^{(l,k)}$ can be expressed in terms of some ``fundamental'' fields (see also \cite{ACDK, DFS1}), 
that satisfy the basic Poisson relations.
Let us  focus on the first two members of the hierarchy, and identify $V^{(1)},\ V^{(2)}$ and the 
corresponding space time difference equations.\\

\noindent {\bf The $V^{(1)}$-operator.} The first non-trivial $V^{(1)}$ of the general  
form is linear and is associated to a discrete time version 
of the transport equation,
\begin{eqnarray}
 V^{(1)}(n,a,\lambda) = \lb \begin{matrix}
		 \lambda +  {\mathbb N}^{(1)}_{na} & {\mathrm X}_{na}  \\
		  {\mathrm Y}_{n-1a}  &  1
	\end{matrix} \rb,  \label{NLSb1f}
\end{eqnarray}
where from the condition $\mbox{det} V^{(1)} = \lambda+ 1$ we obtain
\begin{equation}
{\mathbb N}_{na}^{(1)} = 1 + {\mathrm X}_{na} {\mathrm Y}_{n-1 a}.
\end{equation} 
Due to the fact that $V^{(1)}$ satisfies the quadratic algebra (\ref{Dual2b}) 
and hence (\ref{basica1}), we derive the Poisson relations 
for the fundamental fields (i.e. the time like analogue of (\ref{dn1})):
\begin{equation}
\Big \{ {\mathrm Y}_{na},\ {\mathrm X}_{n-1 b} \Big \}_T= \delta_{ab}.
\end{equation}

Having identified the Lax pair we may now extract the equations of motion associated to  $\Big(L,\  V^{(1)}\Big)$. 
These are linear difference equations in analogy to  the continuous case,
 i.e. they are the discrete analogues of the linear transport equation:
\begin{equation}
F_{n+1a} = F_{na+1}, ~~~F\in \big \{ {\mathrm X},\ {\mathrm Y}\big \}.
\end{equation} $\square$\\

\noindent {\bf The $V^{(2)}$-operator.} The $V^{(2)}$ operator, quadratic in $\lambda$, reads as
\begin{eqnarray}
 V^{(2)}(n,a,\lambda) 
= \lb \begin{matrix}
		\lambda^2+ \lambda {\mathbb N}^{(2)}_{na} + {\mathbb A}_{na}& \lambda {\mathrm X}_{na}+ {\mathbb  B}_{na} \\
		\lambda  {\mathrm Y}_{n-1a} +{\mathbb C}_{na} & {\mathbb D}_{na}
	\end{matrix} \rb, \label{NLSb2f}
\end{eqnarray}
where as in the semi discrete case requiring $\mbox{det}V^{(2)} = \lambda^2 +1$ we obtain the fields
 ${\mathbb N},\  {\mathbb A},\ {\mathbb D}$ expressed in terms of the fundamental fields ${\mathrm X},\ {\mathrm Y}\ {\mathbb B},\  {\mathbb C}$:
\begin{eqnarray}
{\mathbb N}^{(2)}_{na} = {{\mathrm Y}_{n-1a} {\mathbb B}_{na}  +{\mathrm X}_{na}{\mathbb C}_{na} \over 1 +
{\mathrm X}_{na} {\mathrm Y}_{n-1a}}, ~~~
{\mathbb A}_{na} = {1 + {\mathbb B}_{na} {\mathbb C}_{na} \over 1 +{\mathrm X}_{na} {\mathrm Y}_{n-1a}}, ~~~
{\mathbb D}_{na} = 1 +{\mathrm X}_{na} {\mathrm Y}_{n-1a}.  \nonumber\\ \label{FieldsAf}
\end{eqnarray}
Requiring also that $V$ satisfies the time-like Poisson structure  
(\ref{Dual2b}) and hence (\ref{basicb1})-(\ref{basicb3}) we produce the time-like algebra for the fields,  
which reads as (we only write below the fundamental commutators, see also \cite{ACDK}, and Appendix B for the time-like quantum algebra):
\begin{eqnarray}
&&  \Big \lbrace {\mathrm Y}_{n-1a},\ {\mathbb B}_{nb} \Big \rbrace_T  =- \Big \lbrace {\mathrm X}_{na},\ {\mathbb C}_{nb} \Big \rbrace_T  = 
\Big (1 + {\mathrm X}_{na} {\mathrm Y}_{n-1 a}\Big )\delta_{ab}, \nonumber\\
&&  \Big \{{\mathbb B}_{na},\ {\mathbb C}_{nb} \Big\}_T =- \Big( {\mathrm Y}_{n-1a} {\mathbb B}_{na} +{\mathrm X}_{na}{\mathbb C}_{na} \Big) \delta_{ab}.
\label{fund2b}
\end{eqnarray}

We now consider the fully discrete version of the  NLS like model with a Lax pair $\Big (L,\ V^{(2)}\Big )$ given in (\ref{DNLS}) and (\ref{NLSb2f}), 
and we employ the fully discrete zero curvature condition (see all the associated relations in Appendix A).
Then as in the semi-discrete time case studied in the previous subsection we can identify the fields ${\mathbb B},\ {\mathbb C}$
 in terms of ${\mathrm X},\ {\mathrm Y}$ using equations (\ref{EE1}),  (\ref{EE3}) and the definition for ${\mathbb N}^{(2)}$  (\ref{FieldsAf}):
\begin{eqnarray}
&& {\mathbb B}_{na} = {{\mathrm X}_{n+1a} - {\mathbb N}_{na} {\mathrm X}_{na} + {\mathrm X}_{na}^2\big  ({\mathrm Y}_{n-2a}  
-{\mathbb N}_{n-1a+1} {\mathrm Y}_{n-1a} \big )\over  1 - {\mathrm X}_{na} {\mathrm Y}_{n-1a} } \label{xy1} \\
&& {\mathbb C}_{na} = {{\mathrm Y}_{n-1a}^2\big  ({\mathrm X}_{n+1a} - {\mathbb N}_{na} {\mathrm X}_{na}\big ) + {\mathrm Y}_{n-2a}  
-{\mathbb N}_{n-1a+1} {\mathrm Y}_{n-1a} \over  1 - {\mathrm X}_{na} {\mathrm Y}_{n-1a} }  \label{xy2}
\end{eqnarray}
Substituting the above expressions in (\ref{EE2}) and (\ref{EE4}) of the Appendix A
we obtain the quite involved space-time non-linear 
difference equations for the fields ${\mathrm X},\ {\mathrm Y}$
\begin{eqnarray}
&& {\mathbb B}_{n+1a} = {\mathbb N}_{na+1} {\mathbb B}_{na} + 
{\mathrm X}_{na+1}{\mathbb D}_{na} -{\mathbb A}_{n+1a} {\mathrm X}_{na} \label{EE2b}\\
&& {\mathbb C}_{na} = {\mathbb C}_{n+1a} {\mathbb N}_{na+1} +
{\mathbb D}_{n+1a} {\mathrm Y}_{n-1a} - {\mathrm Y}_{na} {\mathbb A}_{na}. \label{EE4b}
\end{eqnarray}
 These equations are the fully discrete analogues of NLS type equations.
Comparing (\ref{xy1})-(\ref{EE4b}) with (\ref{FieldsBp}), (\ref{FieldsCp})  we conclude that  (\ref{xy1})-(\ref{EE4b}) 
are the discrete space analogues of (\ref{FieldsBp}), (\ref{FieldsCp}).
Indeed,  in the continuous space  limit  ${\mathbb N} \to 1 $ and $F_{n+1a}-F_{na} \to \partial_x F_a(x)$,
so equations (\ref{xy1})-(\ref{EE4b})  reduce to (\ref{FieldsBp}), (\ref{FieldsCp}). $\square$

\subsubsection{The fully discrete Ablowitz-Ladik model}

\noindent  We now  examine an alternative version of the fully discrete NLS model associated 
to a trigonometric $r$-matrix. Specifically, we examine the fully discrete version of the AL model.
Indeed, the $r$-matrix for the AL model is a trigonometric one,  a variation of the classical sine-Gordon $r$-matrix \cite{Kulish, FT}:
\begin{equation}
r(\lambda) = {1\over2 \sinh \lambda } \Big (\cosh\lambda  \sum_{j=1}^2 e_{jj} \otimes e_{jj}  +
 \sum_{i \neq j=1}^2 e_{ij} \otimes e_{ji} + \sinh \lambda \sum_{i\neq j=1}^2 (-1)^{j-i} e_{ii} \otimes e_{jj} \Big).\label{rkulish}
\end{equation}
We also recall that the classical Lax operator for the AL model is given by (see e.g. \cite{AL, KakMug, Kono})
\begin{eqnarray}
 L(n, z) = \lb \begin{matrix}
		z & \hat b_n\\
		b_n &  z^{-1}
	\end{matrix} \rb, \label{AL00}
\end{eqnarray}
where $z =e^{\lambda}$ is the multiplicative spectral parameter. The $L$-operator satisfies
 (\ref{Poisson2}), with $r$ being the trigonometric matrix (\ref{rkulish}). 
This leads to the classical  algebra for the fields (see also \cite{KakMug}):
\begin{eqnarray}
&& \Big \{b_n,\ \hat b_m \Big\}_S= \delta_{nm} \big ( 1 - b_n\hat b_m\big ), ~~~
\Big \{b_n,\ b_m  \Big\}_S= \Big\{\hat b_n,\ \hat b_m\Big \}_S=0.  \label{ALPo1}
\end{eqnarray}
The AL model is thus  associated to a deformed harmonic oscillator  classical 
algebra ($q$-bosons at the quantum level \cite{Kulish}).  
Note that $n,\ m$ denote space like indices. Dependence of a 
continuum time-like parameter $t$ or a discrete time dependence characterized by some  time index 
$a$  is implied, but is not explicitly stated for now.  

From the space-like transfer matrix we obtain the following 
space-like conserved quantities, after expanding suitably in powers of $z^{\pm1}$
\begin{equation}
H_S^+ = \sum_{n=1}^{N}\hat \beta_{n+1} \beta_n, ~~~~~H_S^{-} =  \sum_{n=1}^{N}\beta_{n+1} \hat \beta_n.
\end{equation}

Let us also introduce realizations of the time-like algebra (\ref{Dual2b}) with the $r$-matrix given in (\ref{rkulish}).
These realizations will play the role of the  discrete time components of the fully discrete AL Lax pairs: 
\begin{eqnarray}
&& V^-(a, z) = \lb \begin{matrix}
		z & \hat {\mathbb B}_a\\
		{\mathbb B}_a &  -z{\mathbb A}_{a}+ z^{-1}
	\end{matrix} \rb, ~~~ V^+(a, z) = \lb \begin{matrix}
		z-z^{-1}{\mathbb A}_{a} & \hat {\mathbb B}_a\\
		{\mathbb B}_a &  z^{-1}
	\end{matrix} \rb. \label{ALB0}
\end{eqnarray}
Note that here $a,\ b $ denote time indices,  whereas  space dependence
is implied, but is not explicitly stated for now. Requiring that both $V^{\pm}$ satisfy the time-like algebra 
we obtain the associated 
time-like Poisson relations for the fields:
\begin{eqnarray}
&& \Big \{{\mathbb B}_a,\ \hat {\mathbb B}_b \Big\}_T= \delta_{ab} \big ( 1 - {\mathbb B}_a\hat {\mathbb B}_a\big ), ~~~
\Big \{{\mathbb B}_a,\ {\mathbb B}_b  \Big\}_T= \Big\{\hat{\mathbb B}_a,\ \hat {\mathbb B}_b\Big \}_T=0, \nonumber\\
&& \Big\{\hat{\mathbb B}_a,\  {\mathbb A}_b\Big \}_T={\mathbb A}_a\hat {\mathbb B}_a \delta_{ab}, ~~~~ \Big\{{\mathbb B}_a,\  
{\mathbb A}_b\Big \}_T=-{\mathbb A}_a {\mathbb B}_a \delta_{ab}, \label{ALPo2}
\end{eqnarray}
where ${\mathbb A}_a = -1 + \hat {\mathbb B}_a {\mathbb B}_a$ and is compatible with the Poisson structure
above\footnote{${\mathbb A}$ can be defined up to an overall multiplicative constant.}.

From the time-like transfer matrix (\ref{tmono}) we obtain the following time-like conserved quantities corresponding to $V^{\pm}$:
\begin{equation}
H^+_T = \sum_{a=1}^M \Big ( \hat {\mathbb B}_{a+1} {\mathbb B}_a- \hat {\mathbb B}_{a} {\mathbb B}_a\Big ), 
~~~~H^-_T = \sum_{a=1}^M \Big ( \hat {\mathbb B}_{a} {\mathbb B}_{a+1}- \hat {\mathbb B}_{a} {\mathbb B}_a\Big ).
\end{equation}

We shall use suitable Lax pairs to produce  space and time diescretizations of the 
AL model by considering three distinct cases:  \\

\noindent {\bf A.} We first consider the Lax pair $\Big ( L,\ V^-\Big )$:
\begin{equation}
 L(n,a,z) = \lb \begin{matrix}
		z & \hat \beta_{na-1}\\
		\beta_{na} &  z^{-1}
	\end{matrix} \rb,  ~~~~~V^-(n,a,z) = \lb \begin{matrix}
		z & \hat \beta_{n-1a}\\
		\beta_{na} &  -z{\mathbb A}^-_{na}+ z^{-1}
	\end{matrix} \rb
\end{equation}
where ${\mathbb A}^-_{na} = -1 + \beta_{na} \hat \beta_{n-1a}$. From the fully discrete compatibility condition (\ref{full_zero})
the equations of motion arise (from the anti-diagonal entries):
\begin{eqnarray}
&& \hat \beta_{na-1} = \hat \beta_{n-1a} + \hat \beta_{na} - \hat\beta_{na} \beta_{na} \hat \beta_{n-1a} \nonumber\\
&&  \beta_{na+1} =\beta_{n+1a} + \beta_{na} - \beta_{n+1a} \hat \beta_{na}\beta_{na}. \label{A}
\end{eqnarray}
\\
\noindent {\bf B.} We next consider the Lax pair $\Big ( L,\ V^+\Big )$:
\begin{equation}
 L(n,a,z) = \lb \begin{matrix}
		z & \hat \beta_{na-1}\\
		\beta_{na} &  z^{-1}
	\end{matrix} \rb,  ~~~~~V^+(n,a,z) = \lb \begin{matrix}
		z-z^{-1}{\mathbb A}^+_{na} & \hat \beta_{na-1}\\
		\beta_{n-1a+1} &  z^{-1}
	\end{matrix} \rb
\end{equation}
where ${\mathbb A}^+_{na}= -1 +\hat \beta_{na-1} \beta_{n-1a+1}$\footnote{To emphasize the notion of ``ultra-locality'',
and also for our notation to be 
compatible with the rest of the examples, 
we may introduce a new fundamental field
$\gamma_{na-1 } := \beta_{na+1}$, so the fields that appear in $L$ in this case are $\hat \beta_{na-1},$ $\gamma_{na-2}$, and the fields
in $V^+$ are  $\hat \beta_{na-1},$ $\gamma_{n-1na-1}.$} . 
From the fully discrete compatibility condition (\ref{full_zero}) we obtain 
the partial difference equations:
\begin{eqnarray}
&& \hat \beta_{na} = \hat \beta_{n+1a-1} + \hat \beta_{na-1} - \hat \beta_{n+1a-1} \beta_{na+1} \hat\beta_{na-1} \nonumber\\
&&  \beta_{na} =\beta_{n-1a+1} +\beta_{na+1} - \beta_{na+1} \hat \beta_{na-1} \beta_{n-1a+1}. \label{B}
\end{eqnarray}

\begin{rem}{}Interestingly,  by adding equations (\ref{A}) and (\ref{B}) we obtain a space-time discrete analogue
of  an mKdV like equation,  provided that $\beta_{n+1a} \to \hat \beta_{n+1a-1}$:
\begin{eqnarray}
&& \hat \beta_{na} - \hat \beta_{na-1}  =  {1\over 2}\big (1- \hat \beta_{na} \hat\beta_{na-1}\big) 
\big (\hat \beta_{n+1a-1}-  \hat \beta_{n-1 a}\big ) \label{AB}
\end{eqnarray}
\end{rem}
Comparison with  Hirota's lattice KdV reduction \cite{Hirota} would be very interesting, however for such a comparison to be possible
a fully discrete analogue of a Miura-like transformation would be needed. This is a significant open question, 
which however will be addressed elsewhere.

\noindent {\bf C.} Finally we consider the Lax Pair $\Big (L^+,\ V^- \Big)$:
\begin{equation}
 L^+(n,a,z) = \lb \begin{matrix}
		z-z^{-1}\hat {\mathbb A}_{na} & \hat \beta_{na}\\
		\beta_{na-1} &  z^{-1}
	\end{matrix} \rb,~~~~~~V^-(n,a,z) = \lb \begin{matrix}
		z & \hat \beta_{n-1a}\\
		\beta_{na} &  -z{\mathbb A}^-_{na}+ z^{-1},
	\end{matrix} \rb
\end{equation}
where $\hat {\mathbb A}_{na}=-1 +  \hat \beta_{na} \beta_{na-1} $ 
and $ {\mathbb A}^-_{na}= -1 +\beta_{na}\hat \beta_{n-1a}  $, also $L$ is structurally similar to 
$V^+$, but the time and space indices are interchanged, i.e.
$L$ satisfies the space-like algebra and the fields then satisfy the ultra-local Poisson relations for fixed time $a$:
\begin{eqnarray}
&& \Big \{\beta_{na-1},\ \hat \beta_{ma} \Big \}_S = \delta_{nm} \Big ( 1-\beta_{na-1}\hat \beta_{na} \Big ) \nonumber\\
&&  \Big \{\beta_{na-1},\ \beta_{ma-1} \Big \}_S =  \Big \{\hat \beta_{na},\ \hat \beta_{ma} \Big \}_S=0.
\end{eqnarray}
The space time difference equations arising from the fully discrete zero curvature condition read as:
\begin{eqnarray}
&& \hat \beta_{na+1}+ \hat \beta_{n-1a}- \hat \beta_{na}=  \hat \beta_{n-1a} \beta_{na} \hat \beta_{na+1} \nonumber\\
&& \beta_{n+1a}+ \beta_{na-1}- \beta_{na}=  \beta_{n+1a}  \hat \beta_{na}\beta_{na-1}. \label{C}
\end{eqnarray}

Consistency checks have been also performed by comparing the diagonal terms in the 
compatibility condition (\ref{full_zero})  for the three distinct Lax pairs presented  above.

\section{Darboux-dressing formulation $\&$ solutions}

\noindent The most efficient way to derive the continuous time 
components of Lax pairs, i.e. the $V$-operators is the use of the STS formula.
This formula can be derived provided that an associated Poisson structure is available, 
then use of the zero curvature condition
and  Hamilton's equations leads to STS formula \cite{STS}.
However, in the discrete time set up the analogue of the STS
formula is not available, thus alternative ways to
construct the $V$--hierarchy are required.
In the preceding section we were able to construct the 
$V$-operators by requiring that they satisfy the time-like quadratic Poisson structure.
In what follows, mostly as a consistency check on the findings of the previous section,
we implement  the discrete time Darboux-dressing formulation to identify the 
$V$-hierarchy, and confirm the findings of the algebraic approach. 
This process also offers a systematic means to derive solutions of the 
associated integrable  non-linear difference equations as discussed in subsection 3.3.

\subsection{The semi-discrete time NLS hierarchy}
\noindent We first examine the semi-discrete time scenario and consider the Lax pair $\Big (U,\ V\Big)$,
where $U$ is given by $U^{(2)}$ in 
(\ref{Uops}) and the hierarchy of $V$-operators will be derived through the dressing process, 
i.e. we are considering now the space-like description as opposed to the time-like consideration of subsection 2.1.1.  In particular, 
we are going to explicitly derive the first two members 
of the discrete time hierarchy, $V^{(1)}$ and $V^{(2)}$ confirming the algebraic findings of subsection 2.1.
 
Consider the associated  auxiliary linear problem (\ref{ALP}), and let ${\mathbb M}$ be the Darboux transform such that:
\begin{equation}
\Psi(x,a,\lambda) = {\mathbb M}(x,a,\lambda) \hat \Psi(x,a,\lambda),
\end{equation}
where  both $\Psi,\ \hat  \Psi$ are solutions of associated linear problems with Lax pairs 
$\Big (U,\ V\Big)$ and $\Big (\hat U,\ \hat V\Big)$ respectively.
Let us focus on the $x$-part of the linear auxiliary problem to derive the $x$-part of the Darboux-B\"acklund relations:
\begin{equation}
\partial_x {\mathbb M}(x,a,\lambda) = U(x,a,\lambda){\mathbb M}(x,a,\lambda) -
{\mathbb M}(x,a,\lambda) \hat U(x,a,\lambda). \label{xBT}
\end{equation}
We consider here the fundamental Darboux transform for the NLS hierarchy 
(see also recent relevant results for the NLS model and 
generalizations \cite{DoiSkl}, \cite{DFS1})
\begin{equation}
{\mathbb M}(x,a,\lambda) = \lb \begin{matrix}
		\lambda +A_a(x) & B_a(x) \\
		C_a(x) & \lambda +D_a(x)
	\end{matrix} \rb. \label{mm}
\end{equation}
Also, recall that $U$ is given by $U^{(2)}$ in  (\ref{Uops}), and $\hat U=\lb \begin{matrix}
		\lambda  &0 \\
		0 & 0
	\end{matrix} \rb $. Using the fundamental Darboux matrix above and solving the $x$-part of the Darboux-B\"acklund 
transformation (BT) relations (\ref{xBT}) we obtain the following sets of constraints (see also e.g. \cite{DFS1}):
\begin{equation}
 B_a = - \hat u_a, ~~~~~C_a = u_{a-1}, ~~~~~D_a = \Theta -A_a,
\end{equation}
where $\Theta$ is an arbitrary constant, 
and the extra constraints
\begin{eqnarray}
\partial_x \hat u_a =\hat u_a A_a, ~~~~~~\partial_x  u_{a-1} = u_{a-1} A_a,
~~~~~~\partial_x A_a = \hat u_a u_a. \label{xx2}
\end{eqnarray}

We shall also perform the discrete time dressing to obtain the $V$-operators and confirm the expressions for $V$ 
derived algebraically in the previous section. Let us consider the  general form of the $V$-operator 
associated to a certain discrete time
characterized by an index $a$, 
\begin{equation}
V^{(m)}(\lambda) = \lambda^m {\cal D} + \sum_{k=0}^{m-1} \lambda^k {\cal Y}^{(k,m)}, \label{VV}
\end{equation}
 recall ${\cal D}  = diag(1,\ 0)$, and we express ${\mathbb M} = \lambda I +K$,  
where $I$ is the $2\times 2$ identity matrix, and the matrix $K$ reads from (\ref{mm}). Also, $\hat V^{(m)}(\lambda)= \lambda^m{\cal D} +I$.
From the discrete time part of the Darboux relations
\begin{equation}
{\mathbb M}(x,a+1,\lambda) \hat V(x,a,\lambda) = V(x,a,\lambda){\mathbb M}(x,a,\lambda),
\end{equation}
the following recursion relations emerge for the generic object $V^{(m)}$ (\ref{VV}) ($x$ dependence on the expression 
below is always implied, but omitted for brevity):
\begin{eqnarray}
&& {\cal Y}^{(m-1,m)}(a) = K(a+1){\cal D}-{\cal  D} K(a) \nonumber\\
&&  {\cal Y}^{(k-1,m)}(a) = - {\cal Y}^{(k,m)}(a)K(a), ~~~k\in \big \{2, \ldots, m-1 \big \} \nonumber\\
&&  {\cal Y}^{(0,m)}(a)- 1 = -  {\cal Y}^{(1,m)}(a)K(a)\nonumber\\
&& K(a+1) =  {\cal Y}^{(0,m)}(a) K(a). \label{rec1}
\end{eqnarray}
We now focus on the explicit derivation of the first two members of the discrete time hierarchy, $V^{(1)}$ and $V^{(2)}$.

Let us first identify the $V^{(1)}$ operator and the corresponding non-linear ODEs.
The constraints emerging from (\ref{rec1}) associated to $V^{(1)}$ are summarized below:
\begin{eqnarray}
&& {\cal Y}^{(0,1)}_{12}(a) =- B_a, ~~~~~{\cal Y}^{(0,1)}_{21}(a)= C_{a+1}, ~~~~~
{\cal Y}^{(0,1)}_{11}(a)-1 = A_{a+1}- A_a. \label{C1}
\end{eqnarray}
which lead to ${\cal Y}^{(0,1)}_{12}(a)= \hat u_a$, $\ {\cal Y}^{(0,1)}_{21}(a) = u_a$. Also,
\begin{eqnarray}
&& A_{a+1} -A_a = {\cal Y}^{(0,1)}_{11}(a) A_a + \hat u_a  u_{a-1}, ~~~~~A_{a+1}-A_a= u_a \hat u_a \nonumber\\
&& u_a -u_{a-1} = u_a A_a, ~~~~~\hat u_{a+1}- \hat u_a = \hat u_a A_a + \hat u_a^2 u_a. \label{C2}
\end{eqnarray}
Combining  the constraints (\ref{C1}) and (\ref{C2}), and recalling the $x$-part of the Darboux transform (\ref{xx2}) 
we conclude: ${\cal Y}^{(0,1)}_{11}(a)= 1 + u_a \hat u_a$, 
and the nonlinear ODEs
\begin{eqnarray}
&& \hat u_{a+1} - \hat u_a = \hat u_a^2  u_a   + \partial_x \hat u_a \nonumber\\
&& u_{a} - u_{a-1} =-  u_a^2 \hat u_a   + \partial_x  u_a .
\end{eqnarray}
Analogous  expressions were obtained in \cite{DoiSkl}, where the semi-discrete space case was studied. 
In this particular case we observe a simple exchange of the role of space and time. 
The latter equations can be seen as non-linear versions of the transport equation.  We have thus
reproduced expression (\ref{NLSb1}) for $V^{(1)}$ confirming the algebraic approach of the previous section.

We move on to derive $V^{(2)}$ via the dressing formulation. Let us introduce
the following notation compatible with the  
expression  (\ref{NLSb2}) derived in the previous subsection:
\begin{eqnarray}
{\cal Y}^{(0,2)}_{11}(a) ={\mathbb A}_a,  ~~~{\cal Y}^{(0,2)}_{22}(a) = {\mathbb D}_a, ~~~
{\cal Y}^{(0,2)}_{12}(a) ={\mathbb B}_a,  ~~~{\cal Y}^{(0,2)}_{21}(a)  = {\mathbb C}_a.  \nonumber
\end{eqnarray}
and ${\cal Y}^{(1,2)}_{22}(a)=0$, $~{\cal Y}^{(1,2)}_{11}(a) = {\mathbb N}^{(2)}_a$.
Then from equations (\ref{rec1}) for $V^{(2)}$ we obtain for the off diagonal entries
\begin{equation}
{\cal Y}^{(1,2)}_{12}(a) = \hat u_a, ~~~~~~ {\cal Y}^{(1,2)}_{21}(a) = u_a
\end{equation}
as well as the following set  of constraints 
\begin{eqnarray}
&& {\mathbb B}_a = \hat u_a A_a +{\mathbb N}^{(2)}_a \hat u_a, \label{A1}\\
&& {\mathbb B}_a A_a = \hat u_{a+1}- {\mathbb A}_a \hat u_a  \label{A2} \\
&& {\mathbb C}_a = -u_a A_a, \label{A3} \\
&&{\mathbb C}_a A_a = u_a-{\mathbb D}_a u_{a-1}.\label{A4}
\end{eqnarray}
The diagonal entries of (\ref{rec1}) lead to:
\begin{eqnarray}
&& A_{a+1} -A_a = {\mathbb C}_a\hat u_a  + u_a\hat u_a A_a, \label{B1} \\
&&  A_{a+1} - A_a = {\mathbb N}_a^{(2)}  \label{B2} \\
&& A_{a+1} = {\mathbb A}_a A_a +{\mathbb B}_{a}u_{a-1}, \label{B3} \\
&& {\mathbb D}_a=1+u_a \hat u_a, ~~~~{\mathbb A}_a = 1 -{\mathbb N}_a^{(2)}A_a-\hat u_{a} u_{a-1}. \label{B4}
\end{eqnarray}

Combining equations (\ref{B1}), (\ref{B2}) and (\ref{A1}) we conclude the ${\mathbb N}^{(2)}_a$  
as expected  is given by expressions (\ref{FieldsA0}).
Similarly, ${\mathbb D}_a$ given in (\ref{B4}) agrees with expression (\ref{FieldsA0}) from the algebraic viewpoint. Also,
for ${\mathbb A}_a$ we conclude via (\ref{B4}) and (\ref{A3}), (\ref{A4}) and the definition of ${\mathbb D}_a$ (\ref{B4}) 
to the expression given by (\ref{FieldsA0}). 
The dressing process  yields exactly the same expression for $V^{(2)}$ as the algebraic formulation of subsection 2.1 
and this is indeed a strong consistency check. Moreover, the equations of motion derived in subsection 2.2 via 
the zero curvature condition (see also (\ref{FieldsB}), (\ref{FieldsC})), are recovered  via equations (\ref{A1})--(\ref{B4}) and using 
the $x$-part of the Darboux-BT relations ({\ref{xx2}). Indeed, 
expressions (\ref{FieldsB}) are immediately recovered by combining (\ref{A1}), (\ref{A3}) and recalling (\ref{xx2}).

\subsection{The fully discrete time NLS hierarchy}

\noindent We come now to the application of the fully discrete 
Darboux-dressing process in order to construct 
the fully discrete NLS hierarchy.
Let ${\mathbb M}$ be the local Darboux transformation such that 
\begin{equation}
\Psi(n,a,\lambda)= {\mathbb M}(n,a,\lambda) \hat \Psi(n,a,\lambda), \label{gauge}
\end{equation}
where both $\Psi,\ \hat \Psi$ satisfy 
the auxiliary linear problem  (\ref{aux2}) with $\Big (L,\ V\Big )$ and $\Big (\hat L,\ \hat V\Big )$ respectively, then it follows:
\begin{equation}
{\mathbb M}(n+1,a,\lambda)\  \hat L(n,a,\lambda) = L(n,a,\lambda)\ {\mathbb M}(n,a,\lambda) \label{BT10}
\end{equation}
where $L$ is given by (\ref{DNLS}) and  $\hat L$  is in general of the same form, but with fields $\hat X,\ \hat Y$, and here we consider
the simple case where $\hat {\mathrm X}= \hat {\mathrm Y} =0$. Similarly for the discrete time components of the 
Lax pair the transformation (\ref{gauge}) leads to
\begin{equation}
 V(n,a,\lambda)\ {\mathbb M}(n,a,\lambda)=  {\mathbb M}(n,a+1,\lambda)\ \hat V(n,a,\lambda).  \label{BT2}
\end{equation}
We consider for now the fundamental Darboux matrix given in (\ref{mm}),  but ${\cal G}_a(x) \to {\cal G}_{na}$, 
where $ {\cal G} \in \{ A,\ B,\ C,\ D\}$.

From the discrete space part of the Darboux-BT relations (\ref{BT10}) we obtain, 
\begin{eqnarray}
B_{na} =- {\mathrm X}_{na}, ~~~~C_{na} = {\mathrm Y}_{n-1a-1}, \label{FieldsC1}
\end{eqnarray}
We also derive, as expected that $D_{na} = 1 - A_{na}$ and ${\mathbb N}_{na} =1 + X_{na} Y_{na-1}$. 
The discrete space dressing has been 
performed in \cite{DoiSkl}),  and detailed computations can be found there.

Let us first derive $V^{(1)}$, being of the form (\ref{NLSb1f}).  From (\ref{BT2}), we obtain relations 
(\ref{C1})-(\ref{C2}) provided that $u_{a}(x) \to {\mathrm Y}_{n-1a},\ \hat u_a(x) \to {\mathrm X}_{na}$ and
${\mathbb N}_a(x) \to {\mathbb N}_{na}$, conforming also that ${\mathbb N}_{na} = 1 +{\mathrm X}_{na}{\mathrm Y}_{na-1}$. 
Similarly, for the derivation of $V^{(2)}$ (\ref{NLSb2f}) we 
obtain via (\ref{BT2}) equations ({\ref{A1})-(\ref{A4}) and (\ref{B1})-(\ref{B4}), 
 but $u_{a}(x) \to {\mathrm Y}_{n-1a},\  \hat u_a(x) \to {\mathrm X}_{na}$ and
${\cal F}_a(x) \to {\cal F}_{na}$, where $ {\cal F} \in \big \{{\mathbb N}^{(2)},\ 
{\mathbb B},\ {\mathbb  C},\ {\mathbb A},\ {\mathbb D}\big \}$, confirming also equations (\ref{FieldsAf})
coming from the algebraic approach.

\subsection{Solutions}

\noindent 
Having derived the $V$-operators of the discrete time  NLS systems
via the dressing methodology we come now to the derivation of solutions 
of the associated integrable non-linear  difference equations.

Solitonic solutions can be obtained from the fundamental  Darboux matrix as in the continuous 
and the semi-discrete space case (see e.g  \cite{DoiSkl} and references therein). 
In fact, by solving the constraints from the space part of the Darboux transform we obtain such solutions. 
We do not provide the detailed computations here, 
however for a more detailed analysis on the derivation of these expression we refer the interested  reader to \cite{DoiSkl}.
In any case, such expressions will be 
also identified in the subsequent section in a more straightforward manner using a different Darboux matrix, 
which provides not only solitonic,
but generic solutions for both the semi-discrete time scenario and the fully discrete case.
We report below the expressions of the stationary solutions  found in the semi-discrete space case \cite{DoiSkl}, which are also valid
in the fully discrete case:
\begin{enumerate}
\item Solitons of type I
\begin{eqnarray}
{\mathrm X}_n = {\xi^{n-1} ( \xi -1){\mathrm x}_1 \over \xi^{n-1}(\xi -1+ {\mathrm d}_1) -{\mathrm d}_1 }, ~~~~~
 {\mathrm Y}_n = { \xi^{-n}(\xi -1)(1-  {\mathrm a}_1) {\mathrm y}_1 \over  \xi^{-n}(\xi -1 +  {\mathrm a}_1)  - {\mathrm a}_1}, \label{y1}
\end{eqnarray}
where ${\mathrm x}_1,\ {\mathrm y}_1$, ${\mathrm a}_1,\ {\mathrm d}_1$ are constants.
Periodic  boundary conditions are valid for all the associated fields and this can be easily checked 
by inspection provided that $ \xi^{N} =1$. Note that
 in the stationary solutions above the discrete time dependence  is naturally introduced: ${\big (\mathrm X}_n,\ {\mathrm Y}_n \big ) 
\to \big ({\mathrm X}_{na},\ {\mathrm Y}_{na-1}\big )$ and $ \xi^{n} \to  \xi^{n} \zeta^{a}$, where  $\zeta -1 = (\xi-1)^2$, 
(see also next section, where a detailed discussion on the related dispersion relations is presented).
The soliton I solutions $\hat u_a(x),\ u_{a-1}(x)$ for the discrete time and continuum space case studied in 
the proceeding section have the same form as in (\ref{y1}), 
but  $\xi^n \to e^{-kx}$ and the dispersion relation becomes $\zeta = k^2+1$  (see also next section).

\item Solitons of type II
\begin{eqnarray}
&& {\mathrm X}_n = {(\bar \xi -1) {\mathrm x}_1 \over (\bar \xi -1 +\bar \kappa \hat {\mathrm d}_1) \eta ^{-n+1}  - 
\bar \kappa \hat {\mathrm d}_1
\epsilon^{-n+1}}, ~~~~{\mathrm Y}_n = {\eta (\tilde \xi -1)(1-\tilde \kappa \hat {\mathrm a}_1 ) {\mathrm y}_1 
\over (\bar \xi - 1 +\bar \kappa \hat {\mathrm a}_1) \eta ^{n}  - \bar \kappa \hat  {\mathrm a}_1\epsilon^{n}}, \nonumber\label{xy2}
\end{eqnarray}
where $\bar \xi = \epsilon \eta^{-1},\ \bar \kappa = \eta^{-1}$, $~\eta = 1 +c,\ \epsilon = 1-c$, 
and $\tilde \xi = \bar \xi^{-1},\ \tilde \kappa = -\bar \kappa \bar \xi^{-1}$ (see also \cite{DoiSkl}), and ${\mathrm x}_1,\ {\mathrm y}_1$, 
$\hat {\mathrm a}_1,\ \hat {\mathrm d}_1$.
As in the case 1 above the time dependence is easily implemented: ${\big (\mathrm X}_n,\ {\mathrm Y}_n \big ) 
\to \big ({\mathrm X}_{na},\ {\mathrm Y}_{na-1}\big )$ and $\eta^{n}\to \eta^{n}\zeta^a$, $\epsilon^{n}\to 
\epsilon^{n}\hat \zeta^a$, where $\hat \zeta-1 =(\epsilon -1)^{2} ,\ \zeta-1= 
(\eta-1)^{2}$, (see also next subsection). Similarly to case 1 the soliton II solutions $\hat u_a(x),\ u_{a-1}(x)$ 
for the discrete time and continuum space case studied in the proceeding section have the same form as in (\ref{xy2}), 
but  $\eta^n \to e^{-k_1x},\ \epsilon^n \to e^{-k_2x},$ and the dispersion relation becomes 
$\zeta = k_1^2+1,\ \hat \zeta = k_2^2+1$,  (see also next subsection on the issue of dispersion relations).

\end{enumerate}

\noindent Note that 2-soliton solutions can be obtained by repeatedly applying the 
fundamental Darboux and using  Bianchi's permutability theorem.  
Detailed computations and explicit expressions of such solutions can 
be found in \cite{DoiSkl} for the semi-discrete space NLS model.
\\

\noindent{\bf Solutions from the Toda type Darboux.} We consider in what follows 
a different type of Darboux transformation, 
the Toda type Darboux (see also e.g. \cite{DoiSkl}).
We shall employ this transformation to identify generic solutions for both the
 semi-discrete time and the fully discrete NLS systems generalizing the findings of \cite{DoiSkl}.\\

\noindent {\it 1.The semi-discrete time NLS.} Let us first discuss the semi-discrete 
time NLS case and derive solution via the Toda type Darboux 
transformation repeating some of the fundamental computations
of \cite{DoiSkl}.  Recall the $U$-operator of the continuous space and discrete time  Lax pair $\Big(U,\ V\Big)$, 
where $U$ is given by $U^{(2)}$ in  (\ref{Uops}). As was shown in \cite{DoiSkl}
in order to derive general solutions of the non-linear ODEs/PDEs  in the simplest possible way we use the 
Toda type Darboux transform:
\begin{equation}
{\mathbb M}(x, a,\lambda )=  \lb \begin{matrix}
		\lambda +A_a  & B_a\\
		C_{a} &  0
	\end{matrix} \rb. \label{typeI}
\end{equation}
The $x$-part of the Darboux transform gives:
\begin{equation}
\partial_x {\mathbb M}(x,a, \lambda) =  U(x, a,\lambda) {\mathbb M}(x, a,\lambda)- {\mathbb M}(x, a,\lambda) 
U_0(x,a, \lambda), \label{xxdarboux}
\end{equation}
where $U_0$ is  also given by $U^{(2)}$ (\ref{Uops}), but $u_a \to u_a^{(0)},\ \hat u_a \to \hat u_a^{(0)}$.
If $u_a^{(0)} =0 $,
then $\hat u_a^{(0)}$ satisfies the linear equation (we consider the example of the NLS-like equation (\ref{FieldsBp})-(\ref{FieldsCp}))
\begin{equation}
 \hat u_{a+1}^{(0)} -  \hat u_{a}^{(0)}=\partial_x^2  \hat u_{a}^{(0)}. \label{linearn}
\end{equation}
The equation above is nothing but the discrete time version of the heat equation.
The solution of the linear equation  $\hat u_a^{(0)}$ can be expressed as
\begin{equation}
\hat u_0 = \sum_{s=1}^S c_s e^{-k_s x +  \Lambda_s a }, ~~~~\mbox{and/or} ~~~~~
\hat u_0 = \int_{\mathbb R} d\lambda c(\lambda) e^{i\lambda x+  \Lambda_{\lambda} a}\label{fourier2}
\end{equation}
with dispersion relations given as
\begin{equation}
\Lambda_s =\mbox{ln} \big ( k_s^2 +1 \big ), ~~~~~\Lambda_{\lambda}=\mbox{ln} \big ( -\lambda^2 +1 \big ).
\end{equation}

From the Darboux relations (\ref{xxdarboux}) we obtain: $B_a = \hat u_{a}^{(0)},\ C_a = u_{a-1}$
\begin{eqnarray}
\partial_x \hat u_a^{(0)} = - A_a \hat u_a^{(0)}, ~~~~\partial_x u_{a-1}=  A_a  u_{a-1}, 
~~~~~\partial_x A_a =   u_{a-1} \hat u_{a}. \label{basis}
\end{eqnarray}
Solving the equations above leads to:
\begin{equation}
u_{a-1}= {g \over \hat u_a^{(0)}} ~~~~\mbox{and} ~~~~~\hat u_a = -g^{-1}{\hat u_a^{(0)} \partial_x^2(\hat u_a^{(0)}) - 
(\partial_x \hat u_a^{(0)})^2 \over \hat u_a^{(0)}}.
\end{equation}
 Choosing for instance the simple linear solutions: 
$\hat u_0 = c_1  e^{- k_1x + \Lambda_1  a} + c_2$  
or  $\hat u_0 = c_1 e^{- k_1x + \Lambda_1a  }+ c_2 e^{- k_2x + \Lambda_2 a}$ 
we obtain one soliton solutions, respectively:
\begin{eqnarray}
&&  u_{a-1} = {g \over c_1  e^{- k_1x + \Lambda_1  a} + c_2} ~~~~~\mbox{type I soliton}\\
&&  u_{a-1} = {g \over c_1  e^{- k_1x + \Lambda_1  a} + c_2  e^{- k_2x + \Lambda_2 a}} ~~~~\mbox{type II soliton},
\end{eqnarray}
and similarly for $\hat u_a$.\\

\noindent {\it 2. The fully discrete NLS.} We focus now on the derivation of the fully discrete 
NLS solutions by means of the Toda type Darboux. 
\begin{equation}
{\mathbb M}(n+1,a,\lambda)\   L_0(n,a,\lambda) = L(n,a,\lambda)\ {\mathbb M}(n,a,\lambda) \label{BT1}
\end{equation}
where the $L$ operator is give by (\ref{DNLS}) and $L_0$ is given by the same expression as $L$, but with 
${\mathrm X}_{na} \to X_{na}^{(0)}$ and $Y_{na} \to Y_{na}^{(0)}$.
As in the semi discrete case  above we are considering the case where $Y_{na}^{(0)} =0$, then it follows from the set 
of equations of motion for the fields (\ref{EE1})-(\ref{HH4}) that ${\mathrm X}_{na}^{(0)}$ 
satisfy the set of linear difference equations:
\begin{equation}
{\mathrm X}_{n+2a}^{(0)} -2{\mathrm X}_{n+1a}^{(0)}+ {\mathrm X}_{na}^{(0)} =
 {\mathrm X}_{na+1}^{(0)} - {\mathrm X}_{na}^{(0)},
\end{equation}
which is the fully discrete analogue of heat equation.
The solutions of the linear difference equations above are of the generic form
\begin{equation}
{\mathrm X}_{na}^{(0)}  =  \sum_{s=1}^S c_s \xi_s^{n} \zeta_s^{a}, ~~~\mbox{and/or} ~~~
{\mathrm X}_{na}^{(0)}   = \int_{|\xi|=1} d\xi \  c(\xi) \xi^{n} \zeta_{\xi}^{a}, \label{four2}
\end{equation}
and the associated dispersion relations are easily extracted in this setting and read as
\begin{equation}
\zeta_s-1 =( \xi_s -1)^2.\label{disp}
\end{equation}

After solving the set of equations provided by (\ref{BT1}) for (\ref{typeI}) we conclude 
(recall we have set ${\mathrm Y}^{(0)}_{na} =0$, see also \cite{DoiSkl}),
\begin{eqnarray}
&& {\mathrm Y}_{na-1} -{\mathrm Y}_{n-1a-1} = {\mathrm Y}_{na-1}A_{na}, \label{1} \\
&& {\mathrm X}^{(0)}_{n+1a} ({\mathrm X}^{(0)}_{na})^{-1} -1 ={\mathrm X}_{na}{\mathrm Y}_{na-1} -A_{n+1a} \label{2}\\
&& A_{n+1a} -A_{na} ={\mathrm X}_{na} {\mathrm Y}_{na-1}. \label{3}
\end{eqnarray}
Via (\ref{1})--(\ref{3}) we obtain 
\begin{eqnarray}
&& {\mathrm Y}_ {na-1}= \prod_{m=2}^n \big ( 1 -  A_{ma} \big )^{-1} {\mathrm Y}_1 , ~~~~
{\mathrm X}_{na} = (A_{n+1a} -A_{na})  \prod_{m=2}^n \big ( 1 - A_{ma} \big ) {\mathrm Y}_1^{-1}. \nonumber
\end{eqnarray}
Having the solution $ {\mathrm X}^{(0)}_n$ at our disposal we can immediately solve for 
\begin{equation} 
1- A_{na} ={\mathrm X}^{(0)}_{n+1a} ({\mathrm X}^{(0)}_{na})^{-1}\ \Rightarrow\ 
\prod_{m=2}^n \big ( 1 - A_{ma} \big ) = {{\mathrm X}^{(0)}_{n+1a}\over {\mathrm X}^{(0)}_2}, \label{aa}
\end{equation}
and hence obtain the explicit expressions for both fields:
\begin{eqnarray}
&& {\mathrm Y}_{na} ={{\mathrm X}^{(0)}_2 \over {\mathrm X}^{(0)}_{n+1a}} {\mathrm Y}_1, ~~~~~~
{\mathrm X}_ {na} = -({\mathrm X}^{(0)}_2)^{-1} {\mathrm Y}_1^{-1}
{{\mathrm X}^{(0)}_{n+2a} {\mathrm X}^{(0)}_{na} -({\mathrm X}^{(0)}_{n+1a})^2 \over {\mathrm X}^{(0)}_{na}}.\label{y}
\end{eqnarray}
Periodic boundary conditions (${\mathrm X}_{N+1} = {\mathrm X}_1,\ {\mathrm Y}_{N+1} = {\mathrm Y}_1$) are valid provided that 
${\mathrm X}^{(0)}_{N+1} ={\mathrm X}^{(0)}_1$.
Also, $X^{(0)}_2$  and $Y_{1}$ (boundary terms) in the expressions above are treated as constants.
Expressions  (\ref{y}) are general new solutions
of the non-linear partial differential equations for the fully DNLS hierarchy in terms of solutions of the fully discrete heat equation.
We describe below two simple solutions of the type (\ref{y}), which reproduce the  two types 
of discrete solitons.

As in the discrete time case examined in the previous subsection we consider the following simple linear solutions:
\begin{enumerate}
\item We first choose
\begin{eqnarray}
 {\mathrm X}^{(0)}_{na} = c_1 + c_2 \xi^{n} \zeta^a \label{11}
\end{eqnarray}
where recall $ \zeta -1= (\xi-1)^2$,( see also (\ref{disp})). We substitute (\ref{11})  in (\ref{y}), 
and we obtain the discrete analogues of type I solitons:
\begin{eqnarray}
{\mathrm X}_{na}  = -{({\mathrm X}^{(0)}_2)^{-1} {\mathrm Y}_1^{-1}c_1 c_2 (\xi-1)^2\over c_2 + 
c_1 \xi^{-n}\zeta^{-a}}, ~~~~~  {\mathrm Y}_{na}  = {{\mathrm X}^{(0)}_2 {\mathrm Y}_1\over c_1+ 
c_2 \xi^{n+1}\zeta^{a}}. \label{soliton1}
\end{eqnarray}
\item The second simple choice is
 \begin{eqnarray}
{\mathrm X}^{(0)}_n = c_1\eta^{n} \zeta_{\eta}^{a} +  c_2\epsilon^{n}\zeta_{\epsilon}^{a} 
\end{eqnarray}
where $\zeta_{\eta, \epsilon}$ are given by (\ref{disp}).
After substituting the above in (\ref{y}) we obtain the type II discrete solitons:
\begin{eqnarray}
{\mathrm X}_{na}  = -{ ({\mathrm X}^{(0)}_2)^{-1} {\mathrm Y}_1^{-1}c_1 c_2 (\eta - \epsilon)^2\over 
c_1 \epsilon^{-n} \eta_{\epsilon}^{-a} + c_2\eta^{-n}\zeta_{\eta}^{-a}}, ~~~ 
 {\mathrm Y}_n  = {{\mathrm X}^{(0)}_2 {\mathrm Y}_1 \over  c_1 \eta^{n+1} \zeta_{\eta}^a +
c_2 \epsilon^{n}  \zeta_{\epsilon}^a}. \label{soliton2}
\end{eqnarray}
\end{enumerate}
With this we conclude our explicit computation of the two types of discrete soliton solutions for the fully discrete NLS model.
For generic Fourier transforms of the solutions of the linear problem we obtain distinct solutions of the fully discrete NLS.

\section{The two dimensional quantum lattice}

\noindent Our goal now is to generalize the fully discrete description in the quantum case
by constructing the two dimensional quantum lattice.  Out basis for such a construction will be
the fundamental RTT
scheme for deriving quantum algebras (see e.g. \cite{FadTakRes, Korepin}). We first  briefly review this formulation
and then we use it for the construction of  the two dimensional quantum lattice.
For a given $R$-matrix, solution of the Yang-Baxter equation, associated 
quantum algebras emerge from the core relation:
\begin{equation}
R(\lambda_1 -\lambda_2)\ \big ({\cal L}(\lambda_1) \otimes I\big )\ \big (I \otimes {\cal L}(\lambda_2)\big )= 
\big (I \otimes {\cal L}(\lambda_2)\big )\ \big ({\cal L}(\lambda_1) \otimes I\big )\  R(\lambda_1 -\lambda_2) \label{rvv}
\end{equation}
As in the classical frame the ${\cal L}$-operator is the fundamental object and encodes the key algebraic information.

Before we proceed with our construction let us first introduce the
``double quantum space''  notation, which is suitable for the description of the two dimensional quantum lattice. 
Let ${\cal A}_S$ and ${\cal A}_T$ denote
the spatial and temporal quantum algebras respectively,
and let us also distinguish  two types of  ${\cal L}$ operator: space-like Lax operators
$L \in \mbox{End}({\mathbb C}^d) \otimes {\cal A}_{S} \otimes {\cal A}_{T}^{\otimes k}$,
versus time-like operators, i.e. the quantum analogues of $V$-operators, 
$V \in \mbox{End}({\mathbb C}^d) \otimes {\cal A}^{\otimes l}_{S} \otimes {\cal A}^{T}$, both satisfying (\ref{rvv}). 
In the examples we are considering here $k=l=2$.
In the double quantum index notation for $L(n,a)$ ($n$ space index and $a$ times index), 
${\cal A}_{S}$ occupies the $n^{th}$ site in the space-like tensor product,
whereas ${\cal A}_T^{\otimes k}$ occupy the sites $a-k+1$ to $a$ in the time-like tensor product. 
An analogous interpretation holds for $V(n,a)$.

In the space-like description, precisely as in the classical case, we ``freeze'' the time index and we 
construct the one dimensional space monodromy
$T_S(N, 1, a, \lambda) \in  \mbox{End}({\mathbb C}^d) \otimes {\cal A}_S^{\otimes N}\otimes {\cal A}_T^{\otimes k}$ as in (\ref{space1}),
whereas in the time-like description  we freeze space indices and construct the time-like mondromy 
$T_T(n, M,1, \lambda) \in  \mbox{End}({\mathbb C}^d) \otimes {\cal A}_S^{\otimes l}\otimes {\cal A}_T^{\otimes M}$  as in (\ref{tmono}).
Naturally $T_S$  and $T_T$ satisfy (\ref{rvv}) and consequently traces over the auxiliary space lead to
 commuting transfer matrices: ${\mathfrak t}_S \in{\cal A}_S^{\otimes N}\otimes {\cal A}_T^{\otimes k}$
and ${\mathfrak t}_T \in {\cal A}_S^{\otimes l}\otimes {\cal A}_T^{\otimes M}$. 

It is worth noting that in the space transfer matrix the discrete time dependence is considered to be implicit, and similarly  in 
the time transfer matrix the space dependence is implicit.
The term ``quantum spaces'',  albeit  slightly misleading, refers in general 
 to copies of the spatial and temporal quantum algebras (that might be also represented).
The double  quantum index notation for the quantum Lax pair $\Big (L(n,a),\ V(n, a) \Big )$  is also compatible with the classical notation
of section 2. The figures in  pages 11-12 as well as relevant comments on space and time monodromies, and the two dimensional lattice
apply in the quantum case as well. Specifically, the purple and green lines in these figures
correspond as expected to spatial and temporal quantum spaces respectively. A concrete frame  
that describes two dimensional quantum integrable  lattices is provided by the so called 
tetrahedron equation \cite{Zamo, Bazh}. Our construction is more straightforward in the sense that the 
partial  quantum algebras we are interested in are independent of each other, 
and they both emerge from the fundamental relation (\ref{rvv}) as argued above.
In fact, both space and time algebras can be embedded in a bigger algebra, 
which however simply decomposes into two independent parts ruled by (\ref{rvv}).

We next examine the quantum versions of the two main examples considered in the classical case, 
i.e. the fully discrete NLS and AL models.

$ $

\noindent {\it 1. The discrete NLS model.} We first examine the quantum DNLS system.
Inspired by the classical expressions we consider the generic algebraic  objects of the form 
\begin{equation}
{\cal L}^{(m)}(\lambda)= \sum_{k=0}^{m}\lambda^k {\cal Y}^{(k,m)}, \label{qstr}
\end{equation} where ${\cal Y}^{(m,m)}=  diag (1,\ 0)$.\\

\noindent $\bullet $ {\bf The basic assumptions}
\begin{enumerate}

\item The ${\cal L}^{(m)}$-operators  satisfy the quantum algebra (\ref{rvv}),
where $R(\lambda) = \lambda + {\cal P}$ is the Yangian $R$-matrix, and recall ${\cal P} =\sum_{i,j=1}^de_{ij} 
\otimes e_{ji}$ is the permutation operator for the general $\mathfrak{gl}_d$ case. \\

\item We require the existence of an inverse:
\begin{equation}
{\cal L}^{(m)}(\lambda)\bar {\cal L}^{(m)}(-\lambda) = f^{(m)}(\lambda)I \label{cond1}
\end{equation}
where $f^{(m)}(\lambda) = \lambda^m +\sum_{k=0}^{m-1}a_k\lambda^k$, and
we define
\begin{equation}
\bar {\cal L}(\lambda) = \big ({\cal U}\otimes \mbox{id}\big ) {\cal L}^{t_a}(-\lambda-1) 
\big ({\cal U}\otimes \mbox{id}\big ), \label{Lmatrix}
\end{equation}
${\cal U} =antidiag(i,\ -i)$ and $^{t_a}$ denotes transposition with respect to the two dimensional in our case
($d$ dimensional  in general),``auxiliary'' space.
Specifically, let ${\cal L}(\lambda) = \sum_{i,j} e_{ij} \otimes {\cal L}_{ij}(\lambda)$ then 
${\cal L}^{t_a}(\lambda) = \sum_{i,j} e_{ji} \otimes {\cal L}_{ij}(\lambda)$.
\end{enumerate}
The condition (\ref{cond1}) is equivalent to the requirement that the quantum determinant 
of ${\cal L}^{(m)}$ is proportional to the identity, in analogy to the classical case. The problem 
thus reduces into deriving realizations of the quantum algebra of the form (\ref{rvv}) 
subject to the constraint (\ref{cond1}).

Let us focus on the first two elements of the algebraic hierarchy ${\cal L}^{(1)}$ and ${\cal L}^{(2)}$ 
assuming that they provide realizations of the
quantum algebra (\ref{rvv}).  Let us express ${\cal L}^{(1)},\ {\cal L}^{(2)}$ as follows
\begin{eqnarray}
&& {\cal L}^{(1)}(\lambda) 
= \lb \begin{matrix}
		\lambda+ {\mathbb N}& {\mathrm X} \\
		 {\mathrm Y} & 1
	\end{matrix} \rb, ~~~~{\cal L}^{(2)}(\lambda) 
= \lb \begin{matrix}
		\lambda^2+ \lambda {\mathbb N}^{(2)} + {\mathbb A}& \lambda {\mathrm X}+ {\mathbb B} \\
		\lambda  {\mathrm Y} + {\mathbb C} & {\mathbb D}
	\end{matrix} \rb. \label{NLSb2g}
\end{eqnarray}
\\
\noindent {\bf The ${\cal L}^{(1)}$- operator.} For $m=1$ and $f^{(1)}= \lambda+1$ we recover the DNLS model \cite{KunduRagnisco, Sklyanin}. 
Indeed, solving  condition (\ref{cond1}) we conclude that ${\mathbb N}= 1 +{\mathrm X}{\mathrm Y }$ and due to the fact that ${\cal L}^{(1)}$ 
satisfies the quantum algebra (\ref{rvv}) we obtain the familiar canonical relations for the fields:
$\big [{\mathrm X},\ {\mathrm Y} \big ] = 1,$
and the extra  relations $\big [ {\mathrm  X},\ {\mathbb  N}\big ] ={\mathrm  X}$ compatible with the definition of 
${\mathbb N}$ from (\ref{cond1}). A familiar representation of the canonical fields is given in terms of differential operators
as ${\mathrm X} \mapsto x \xi,~~{\mathrm Y} \mapsto x^{-1}\partial_{\xi},$
where $x$ commutes with both $\xi,\ \partial \xi$. $\square$
\\

\noindent {\bf The ${\cal L}^{(2)}$-operator.} For $m=2$ and $f(\lambda)  = \lambda^2 +a_1\lambda + a_0$, then condition (\ref{cond1}) 
gives rise to the following identities (we choose $a_1=1$)\footnote{There is a freedom on the derivation of the fields up to constant 
and/or a shift depending on the choice of the constants $a_k$.}
\begin{equation}
{\mathbb D} = 1 +{\mathrm X} {\mathrm Y}, ~~~~
{\mathbb N}^{(2)} = \Big ({\mathrm X} {\mathbb  C} + {\mathbb B} {\mathrm Y}\Big ){\mathbb  D}^{-1} \label{qq1}
\end{equation}
and also
\begin{equation}
{\mathbb A} = {\mathbb D}^{-1} \Big ( a_0 -{\mathrm  X} {\mathbb C} + {\mathbb B} {\mathbb C} \Big )
\label{qq2}
\end{equation}
The above expressions  (\ref{qq1}) and (\ref{qq2}) are the quantum analogues of (\ref{fund2b}). 
The algebraic relations between the fundamental fields 
are dictated by (\ref{rvv}) and are given as follows, we first give the exchange relations among the fundamental fields:
\begin{eqnarray}
&& \big[ {\mathrm  X},\ {\mathrm  Y}\big ] =0, ~~~\big[ {\mathbb  B},\ {\mathbb  C}\big ] =  {\mathbb  N}^{(2)}{\mathbb D}, ~~~~
\big[ {\mathrm X},\ {\mathbb  C}\big ] = {\mathbb D}, ~~~\big[ {\mathrm Y},\ {\mathbb B}\big ] =- {\mathbb  D} \label{qtime}
\end{eqnarray}
 All the exchange relations among the various fields emerging from (\ref{rvv}) 
are presented in Appendix B.
The semi-classical limit of the quantum algebraic relations above indeed lead to 
the Poisson relations (\ref{basicb1})-(\ref{basicb3}) and  (\ref{fund2b}),
provided that $-\big [\ ,\ \big ] \to \big \{\ ,\ \big\}$

A representation of the algebra (\ref{qtime}) in terms of differential operators is given below
\begin{equation}
{\mathrm X} \mapsto fx, ~~~{\mathrm Y} \mapsto g y, ~~~{\mathbb B} \mapsto g^{-1}\big  (1 + fg x y\big ) \partial_y, 
~~~{\mathbb C} \mapsto - f^{-1} \big (1 + fg x y\big ) \partial_x, \label{rept}
\end{equation}
where $f,\ g$ commute with each other and also commute with $x,\ y,\ \partial_x\ \partial_y$.
Also, as is well known typical realizations of the algebra (\ref{qtime}) are obtained as tensor products of the algebra, i.e. we  define 
${\cal L}^{(2)}(\{j\}, \lambda)= {\cal L}^{(1)}(j+1, \lambda) {\cal L}^{(1)}(j,\lambda) $, 
where here we use the index notation and $j$  can be either space or time index. $\square$
\\

Given the form of the ${\cal L}$--operators we derived above we can now identify the quantum 
Lax pair for the discrete space-time NLS system expressed in the double index notation. The space component is given by
$~{\cal L}^{(1)} \to L(n, a):  ~{\mathrm X} \to {\mathrm X}_{na}, ~{\mathrm Y} \to {\mathrm Y}_{na-1},$
and the time component:
${\cal L}^{(2)} \to V^{(2)}(n, a):  ~{\mathrm X} \to {\mathrm X}_{na},~
{\mathrm Y} \to {\mathrm Y}_{n-1a}, ~~{\cal F} \to {\cal F}_{na},$
where ${\cal F} \in \big \{ {\mathbb B},\ {\mathbb C},\ {\mathbb N}^{(2)},\ {\mathbb A},\  {\mathbb D} \big \}$. 

The elements of the  temporal quantum algebra, for a fixed $n$, can be expressed in the double quantum index notation as (\ref{rept})
\begin{eqnarray}
&& {\mathrm X}_{na} \mapsto f_n x_a, ~~~~{\mathrm Y}_{n-1a} \mapsto g_{n-1} y_a, \nonumber\\
&& {\mathbb B}_{na} \mapsto g_{n-1}^{-1} \big  (1+ f_n g_{n-1} x_a y_a\big ) \partial_{y_a}, \nonumber\\
&& {\mathbb C}_{na} \mapsto - f_{n}^{-1} \big (1+ f_n g_{n-1} x_a y_a\big ) \partial_{x_a},
\end{eqnarray} 
where in the expressions above a representation for $f_n, \ g_n$, compatible with the space like algebra, can be given as 
$f_n \mapsto \xi_n,\ g_n \mapsto \partial_{\xi_n}$.
\\

\noindent {\it 2. The quantum AL model.} Let us now focus on the case of a trigonometric $R$- matrix 
and the quantum versions of the AL model.
The quantum AL model. Consider now various solutions of the RTT relations (\ref{rvv})
 in the case we choose the trigonometric matrix \cite{Kulish}:
\begin{equation}
R(\lambda) = a(\lambda)  \sum_{j=1}^2 e_{jj} \otimes e_{jj}  +
c\sum_{i \neq j=1}^2 e_{ij} \otimes e_{ji} +b(\lambda)\sum_{i\neq j=1}^2 q^{sgn (j-i)} e_{ii} \otimes e_{jj}, \label{kul}
\end{equation}
where $q=e^{\mu}$ and $a(\lambda)=\sinh{\lambda + \mu} ,\ b(\lambda)=\sinh \lambda,\ c = \sinh{\mu}.$

We consider below the quantum analogues of the three distinct cases 
discussed in the classical case:
\begin{equation}
 {\cal L}(z) = \lb \begin{matrix}
		z & \hat b\\
		b &  z^{-1}
	\end{matrix} \rb,\label{ll1}
\end{equation}
and the associated quantum algebra (\ref{rvv}) is given as (see also \cite{Kulish}) (recall $z=e^{\lambda}$)
\begin{equation}
q\hat b b - q^{-1}b \hat b = q-q^{-1}\label{qua1}
\end{equation}
We also consider the following ${\cal L}$-operators, solutions of (\ref{rvv})
\begin{equation}
{\cal L}^-(z) = \lb \begin{matrix}
		z & \hat {\mathbb B} \\
		{\mathbb B} &  -z{\mathbb A}+ z^{-1}
	\end{matrix} \rb,~~~{\cal L}^+(z) = \lb \begin{matrix}
		z-z^{-1}{\mathbb A} & \hat {\mathbb B}\\
		{\mathbb B} &  z^{-1}
	\end{matrix} \rb, \label{vv1}
\end{equation}
where ${\mathbb A} =- 1 +\hat {\mathbb B} {\mathbb B}$ (defined up to an overall multiplicative constant).
The corresponding quantum algebra:
\begin{eqnarray}
&& q\hat {\mathbb B} {\mathbb B} - q^{-1}{\mathbb B} \hat {\mathbb B} = q-q^{-1}, ~~~\hat {\mathbb B} {\mathbb A} = 
q^{-2} {\mathbb A}\hat {\mathbb B}, ~~~{\mathbb B} {\mathbb A} = 
q^{2} {\mathbb A}{\mathbb B} \label{qua2}
\end{eqnarray}
The semi-classical limit of the quantum algebraic relations (\ref{qua1}), (\ref{qua2}) 
lead to the  Poisson relations (\ref{ALPo1}), (\ref{ALPo2}),
provided that $~-{1\over 2\mu}\big [\ ,\ \big ] \to \big \{\ ,\ \big\}$.\\

Representations of the algebras (\ref{qua1}), (\ref{qua2}) are provided as follows 
(see also \cite{Korepin, DoikouXXZ} and relevant references therein).
Let ${\mathbb X},\ {\mathbb Y}$: $~{\mathbb X} {\mathbb Y} = q^2 {\mathbb Y} {\mathbb X}$, then
\begin{equation}
\hat b := (q \xi{\mathbb X} +1){\mathbb Y}\zeta, ~~~b := {\mathbb Y}^{-1}\zeta^{-1}
\end{equation}
where $\xi,\zeta$ commute with ${\mathbb X},\ {\mathbb Y}$ and they commute with each other,
similarly for $\hat {\mathbb B},\ {\mathbb B}$. Typical realizations of the elements ${\mathbb X},\ {\mathbb Y}$ are given as
~${\mathbb X} := e^{\hat {\mathrm x}}, ~~{\mathbb Y} := e^{\hat{ \mathrm y}}$
provided that $\big [ \hat {\mathrm x},\ \hat {\mathrm y}\big ] =2\mu$ ($q=e^{\mu}$). 
${\mathbb X},\ {\mathbb Y}$  can be represented in terms
of differential operators: $ \hat {\mathrm x} \mapsto-2\mu x,\  \hat {\mathrm y} \mapsto \partial_x$, or by matrices, for example
${\mathbb X} \mapsto \sum_{k=1}^p q^{-2k} e_{kk}$ and  ${\mathbb Y} \mapsto \sum_{k=1}^{p-1} e_{kk+1}+ e_{p1}$. 
The latter $p$ dimensional representation is called 
the cyclic representation and is valid for $\mu = {\pi \over p}$.

The operators ${\cal L},\ {\cal L}^{\pm}$ will be now  used for realizing the quantum discrete AL model.
Below, we express the quantum Lax pairs in the double quantum index notation.\\
\noindent {\bf A.} We first consider the Lax pair $\Big ( L,\ V^-\Big )$:
${\cal L} \to L(n,a): ~\hat b \to \hat \beta_{na-1}, ~b \to \beta_{na}$ and
$~{\cal L}^- \to V^-(n,a): ~\hat {\mathbb B} \to \hat \beta_{n-1a}, ~{\mathbb B} \to \beta_{na}.$\\
\noindent {\bf B.} We also consider the Lax pair $\Big ( L,\ V^+\Big )$:
${\cal L} \to L(n,a): ~\hat b \to \hat \beta_{na-1}, ~b \to \gamma_{na-2}$ and
$~{\cal L}^+ \to V^+(n,a): ~\hat {\mathbb B} \to \hat \beta_{na-1}, ~{\mathbb B} \to \gamma_{n-1a-1}.$
\\
\noindent {\bf C.} Finally we consider the Lax Pair $\Big (L^+,\ V^- \Big)$:
${\cal L}^+ \to L^+(n,a): ~\hat {\mathbb B} \to \hat \beta_{na}, ~{\mathbb  B} \to \beta_{na-1}$
and $V^{-}(n,a)$ is defined as in case A.

When defining the object ${\mathbb A} = -1 +\hat {\mathbb B} {\mathbb B}$  appearing in ${\cal L}^{\pm}$
we consistently keep a specific order for the fields involved, which of course is irrelevant at the classical level.
Similarly, the order in the non-linear terms in the partial difference equations 
(\ref{A}), (\ref{B}), (\ref{C}) is important in the quantum case. Our quantum description is  also compatible with the notion of the 
quantum auxiliary linear problem and the quantum Darboux-B\"acklund transformation as discussed in \cite{DoikouFindlay1, Korff}.
The various quantum equations of motion of (\ref{A}), (\ref{B}) and (\ref{C})
are as expected  precisely of the form of the quantum Darboux-B\"acklund relations appearing in \cite{DoikouFindlay1, Korff}, 
due to the form of the discrete zero curvature condition (\ref{full_zero}) (cf. (\ref{BT10})).
Moreover, the algebraic content of the quantum Darboux matrix as suggested in \cite{DoikouFindlay1} is fully justified 
by the existence of space-time quantum algebras, and in particular the fact that the $V$-operator, 
which plays the role of the Darboux matrix, satisfies the temporal quantum algebra.

\begin{rem}{} The trigonometric $R$-matrix  (\ref{kul})
as well as the various expressions for the ${\cal L}$-operators can 
be associated to the more familiar ${\mathfrak U}_q(\mathfrak{sl}_2)$
$R$-matrix \cite{Jimbo} (see for instance the use of the various versions in 
\cite{Kulish, DoikouFindlay1, Korff}) via suitable transformations.
\end{rem}

\noindent Indeed, let ${\cal L}(\lambda) =\big (G^{-1} \otimes V^{-1} \big ) \hat {\cal L}(\lambda)  \big ( G \otimes\mbox{id}\big )$ and 
$R(\lambda) = \big (G \otimes G^{-1} \big )\hat R(\lambda)\big ( G \otimes G^{-1}\big )$, where $ G = diag(q^{{1\over 4}},\  q^{-{1\over 4}})$ 
and $\hat R$ is   
the XXZ (or sine-Gordon) $R$ matrix is given as \cite{Jimbo}
\begin{equation}
\hat R(\lambda) = a(\lambda)  \sum_{j=1}^2 e_{jj} \otimes e_{jj}  +
c\sum_{i \neq j=1}^2 e_{ij} \otimes e_{ji} +b(\lambda)\sum_{i\neq j=1}^2 e_{ii} \otimes e_{jj}, \label{kul2}
\end{equation}
where recall, $q=e^{\mu}$ and $a(\lambda)=\sinh{\lambda + \mu} ,\ b(\lambda)=\sinh \lambda,\ c = \sinh{\mu}.$ 
Also, the algebraic object 
$V$ is such that,  ${\mathbb A} = V^{-2}$, equivalently,  
$\hat {\mathbb B} V =q V \hat {\mathbb B}$ and  ${\mathbb B} V =q^{-1} V {\mathbb B}$.
Also, $det_q \hat {\cal L} \propto  \mbox{id}$ or equivalently (\ref{cond1}) is valid for $\hat {\cal L}$.
It can be shown by direct computation for (\ref{ll1}) and (\ref{vv1}) 
that  $\big (G^{-1} \otimes V^{-1} \big ) \hat {\cal L}(\lambda)  \big ( G \otimes\mbox{id}\big )
=\big (G \otimes \mbox{id} ) \hat {\cal L}(\lambda)  \big ( G^{-1} \otimes V^{-1}\big )$. $\hat {\cal  L}$ 
satisfies relation (\ref{rvv}) with
the $\hat R$-matrix (\ref{kul2}). $\square$

\begin{rem}{} The  ${\cal L}$ and $\hat {\cal L}$ operators 
are associated to the same quantum algebra, however they provide distinct co-products.
\end{rem}
\noindent Let us use ${\cal L}^{\pm}$ as our examples to illustrate this. Let
\begin{equation}
\hat {\cal L}^-(z) = \lb \begin{matrix}
		zV & \hat {\mathbb C} \\
		{\mathbb C} &  -zV^{-1}+ z^{-1}V
	\end{matrix} \rb,~~~\hat {\cal L}^+(z) = \lb \begin{matrix}
		zV-z^{-1}V^{-1} & \hat {\mathbb C}\\
		{\mathbb C} &  z^{-1}V
	\end{matrix} \rb, \label{vv2}
\end{equation}
where ${\mathbb C} = q^{-{1\over 2}} V{\mathbb B},\ \hat {\mathbb C} = q^{1\over 2} V\hat {\mathbb B}$.
We also multiply ${\hat {\cal L}}$ and ${\cal L}$ with $\sigma^z$ (the diagonal Pauli matrix), 
($(\sigma^z \otimes \sigma^z) R (\sigma^z \otimes \sigma^z) =R$,  same for $\hat R$).
The quantum algebras emerging form (\ref{rvv}) are Hopf algebras equipped with a co-product   
$\Delta({\cal L}(\lambda)) ={\cal  L}(2, \lambda){\cal L}(1, \lambda )$, where $1,\ 2$ are quantum space indices.
Then from ${\cal L}^-$ we obtain (recall  ${\mathbb A} = V^{-2}$)
\begin{equation}
\Delta({\mathbb B}) = {\mathbb B} \otimes \mbox{id}, ~~~~
\Delta(\hat {\mathbb B}) = \hat {\mathbb B} \otimes \mbox{id} +V^{-2} \otimes \hat {\mathbb B},
\end{equation}
whereas from $\hat {\cal L}^-$:
\begin{equation}
\Delta({\mathbb C}) = {\mathbb C} \otimes V, ~~~~
\Delta(\hat {\mathbb C}) = \hat {\mathbb C} \otimes V +V^{-1} \otimes \hat {\mathbb C}
\end{equation}
Similarly, from ${\cal L}^+$ ($\hat {\cal L}^+$) with ${\mathbb B},\ \hat {\mathbb B}$ (${\mathbb C},\ \hat {\mathbb C}$)
being interchanged in the co products above. $\square$

We have only considered here periodic  boundary conditions at both classical and quantum level.
The significant point then is the  implementation of 
integrable space and time integrable  boundary conditions \cite{Sklyanin2, AvanDoikou1, DFS1} in 
the discrete systems, and the effect of these boundary conditions on the
behavior of the solutions. As a final remark we note that
although the fully discrete case represented various technical and conceptual difficulties,
we were able to achieve the consistent simultaneous discretizations of both space and time directions in such 
a way that integrabilty was ensured, based on the concurrent existence of temporal and spatial classical and quantum algebras.

\subsection*{Acknowledgments}
\noindent AD acknowledges support from the  EPSRC research grant  EP/R009465/1.

\appendix

\section{Discrete time NLS equations: consistency}

\noindent {\it 1. Semi-discrete time NLS.} We obtain the following set of constraints from the zero curvature condition, 
by focusing first on the off diagonal elements:
\begin{eqnarray}
&& {\mathbb B}_a =\partial_x \hat u_a + {\mathbb N}^{(2)}_a \hat u_a \label{f1}\\
&& \partial_x {\mathbb B}_a = \hat u_{a+1} {\mathbb D}_a -{\mathbb A}_{a} \hat u_a \label{f2}\\
&& {\mathbb C}_a= - \partial_x u_a + u_a {\mathbb N}_a^{(2)} \label{f3}\\
&& \partial_x {\mathbb C}_a = u_a {\mathbb A}_a - {\mathbb D}_a u_{a-1} \label{f4}
\end{eqnarray}
Dependence of the fields on $x$ is in all the always implied, but omitted for brevity.
Equations (\ref{f1}), (\ref{f3}), via the definition of ${\mathbb N}^{(2)}$ (\ref{FieldsA0}), lead to
\begin{equation}
{\mathbb B}_a = {\partial_x\hat u_a - \hat u_a^2 \partial_x u_a \over 1-u_a \hat u_a}, ~~~~~
{\mathbb C}_a = { u_a^2 \partial_x \hat u_a - \partial_x u_a \over 1-u_a \hat u_a}.
\label{FieldsB}
\end{equation}
Also, from the zero curvature condition we obtain the following constraints
\begin{equation}
\partial_x {\mathbb  B}_a = \hat u_{a+1}{\mathbb D}_a - {\mathbb A}_a \hat u_a, 
~~~~\partial_x {\mathbb C}_a  = u_a {\mathbb A}_a 
- {\mathbb D}_a u_{a-1}. \label{FieldsC}
\end{equation}
Given that that ${\mathbb A},\ {\mathbb D},\ {\mathbb N}^{(2)}$ (\ref{FieldsA0}), 
and ${\mathbb B},\ {\mathbb C}$ (\ref{FieldsB})
are expressed in terms of the fundamental fields $u_a,\hat u_a$ and their 
$x$-derivatives, equations (\ref{FieldsC})
are the equations of motion for the fundamental fields
fields $u_a,\ \hat u_a$. 

Consistency checks are also performed for the diagonal  entries of the matrix from the zero curvature condition:
\begin{eqnarray}
&&  \partial_x {\mathbb D}_a = u_a {\mathbb B}_a - {\mathbb C}_a \hat u_a \label{dg1}\\
&&  \partial_{x} {\mathbb N}^{(2)} = \hat u_{a+1} u_a - \hat u_a u_{a-1} \label{dg2}\\
&&  \partial_x {\mathbb A}_a= \hat u_{a+1} {\mathbb C}_a - {\mathbb B}_a u_{a-1}. \label{dg3}
\end{eqnarray}
Indeed, using the fundamental equations (\ref{f1})-(\ref{f2}) and the definitions (\ref{FieldsA0}) the 
above equations are confirmed.\\

\noindent {\it 2.  Fully discrete  NLS.}  Having at our disposal the Lax pair 
we can now write down the set of equations merging for the fully discrete 
version of the zero curvature condition (\ref{full_zero})
\begin{eqnarray}
&& {\mathbb B}_{na} =  {\mathrm X}_{n+1a}  + \big ({\mathbb N}_{n+1a}^{(2)} -  
{\mathbb N}_{na+1} \big ) {\mathrm X}_{na} \label{EE1}\\
&& {\mathbb  B}_{n+1a} = {\mathbb N}_{na+1} {\mathbb B}_{na} + {\mathrm X}_{na+1}{\mathbb D}_{na} -
{\mathbb A}_{n+1a} {\mathrm X}_{na} \label{EE2}\\
&& {\mathbb C}_{n+1a} = {\mathrm Y}_{n-1a} - {\mathrm Y}_{na} 
\big (  {\mathbb N}_{na}- {\mathbb N}_{na}^{(2)}\big ) \label{EE3}\\
&& {\mathbb C}_{na} = {\mathbb C}_{n+1a} {\mathbb N}_{na+1} +
{\mathbb D}_{n+1a} {\mathrm Y}_{n-1a} - {\mathrm Y}_{na} {\mathbb A}_{na}\label{EE4}
\end{eqnarray}
The first equations above come form the off diagonal elements of the zero curvature condition. 
Equations (\ref{EE1})-(\ref{EE4}) together with (\ref{FieldsAf})  provide the fully discrete analogues of 
 the equations of motion (\ref{f1})-(\ref{f4}). 

The diagonal entries provide extra consistency constraints
\begin{eqnarray}
&& {\mathbb N}_{n+1a}^{(2)}- {\mathbb N}_{na+1} = {\mathbb N}_{na}^{(2)}- {\mathbb N}_{na}  \label{HH1}\\
&& {\mathbb D}_{n+1a} -{\mathbb D}_{na} = {\mathrm Y}_{na} {\mathbb B}_{na} - {\mathbb C}_{n+1a} {\mathrm X}_{na} \label{HH2}\\
&& {\mathbb A}_{n+1} - {\mathbb A}_n = {\mathbb N}_{na+1} {\mathbb N}_{na}^{(2)} - {\mathbb N}_{n+1 a}^{(2)} {\mathbb N}_{na} 
+{\mathrm X}_{na+1} {\mathrm Y}_{n-1a}-{\mathrm X}_{n+1a} {\mathrm Y}_{na-1} \label{H3}\\
&& {\mathbb A}_{n+1a}{\mathbb N}_{na}-  {\mathbb A}_{na}{\mathbb N}_{na+1}= 
{\mathrm X}_{na+1} {\mathbb C}_{na} - {\mathbb B}_{n+1a} {\mathrm Y}_{na-1}. \label{HH4}
\end{eqnarray} 
Equations  (\ref{HH2})-(\ref{HH4}) are the discrete space analogues  of the semi-discrete case (\ref{dg1})-(\ref{dg3}). 
However in the fully discrete case an extra constraint arising and corresponds to equation (\ref{HH1}).

\section{Algebraic relations $\&$ compatibility}

\noindent By means of (\ref{rvv})  
we obtain the following algebraic relations involving all the fields (\ref{qq1}), (\ref{qq2}):
\begin{eqnarray}
&& \big [ {\mathbb C},\ {\mathbb D} \big ] = - {\mathrm Y} {\mathbb N}^{(2)}, ~~~  
\big [ {\mathbb B},\ {\mathbb D} \big ] =  {\mathbb N}^{(2)}{\mathrm X}, 
~~~~\big[ {\mathrm X},\ {\mathbb D}\big]=\big[ {\mathrm Y},\ {\mathbb D}\big]=0\\
&& \big [ {\mathbb N}^{(2)},\ {\mathbb B}\big ] = - {\mathbb B}, ~~~~ \big [ {\mathbb N}^{(2)},\ {\mathbb C}\big ] = {\mathbb C},~~~
\big [{\mathbb N}^{(2)},\  {\mathbb A} \big ] = \big [{\mathbb N}^{(2)},\  {\mathbb D} \big ] = 0\\
&& \big[ {\mathrm X},\  {\mathbb  A} \big] = {\mathbb B}, ~~~~ \big[ {\mathrm Y},\  {\mathbb A} \big] = -{\mathbb C}, 
~~~~\big[ {\mathrm X},\  {\mathbb N}^{(2)} \big] = {\mathrm X}, ~~~~\big[ {\mathrm Y},\  {\mathbb N}^{(2)} \big] = -{\mathrm Y}\\
&& \big[ {\mathbb  A},\  {\mathbb B} \big] ={\mathbb A} {\mathrm X} - {\mathbb N}^{(2)} {\mathbb B},~~~~
\big[ {\mathbb A},\  {\mathbb  C} \big] =
{\mathbb C} {\mathbb N}^{(2)} - {\mathrm Y} {\mathbb A}\\
&&  \big[ {\mathbb A},\  {\mathbb D} \big] ={\mathbb  C} {\mathrm X} - {\mathrm Y} {\mathbb B},~~~~ 
\big[ {\mathrm X},\  {\mathbb D} \big] =\big[ {\mathrm Y},\  {\mathbb D} \big] =0.
\end{eqnarray}
All the relations above are compatible with the fields as defined in (\ref{qq1}), (\ref{qq2}) and (\ref{qtime}).


\begin{thebibliography}{99}

\bibitem{AKNS1}
 M.J. Ablowitz, D. J. Kaup, A. C. Newell and H. Segur, Phys. Rev. Lett. 31 (1973)
125.

\bibitem{AL}
M.J. Ablowitz and J.F. Ladik, J. Math. Phys. 16 (1975) 598.

\bibitem{Ablo} 
M.J. Ablowitz, B. Prinari and A.D. Trubatch, 
{\it Discrete and Continuous Nonlinear Schr\"odinger Systems}, 
London Mathematical Society Lecture Notes, Vol. 302, (2004), Cambridge University Press.

\bibitem{ADP} 
P. Adamopoulou, A. Doikou and G. Papamikos, Nucl. Phys. B918 (2017) 91.

\bibitem{ABS}
V. Adler, A. Bobenko and Yu. Suris, Comm. Math. Phys. 233 (3) (2003) 513.

\bibitem{AvanDoikou1} 
J. Avan, A. Doikou, Nucl. Phys. B800 (2008) 591.

\bibitem{ACDK} 
J. Avan, V. Caudrelier, A. Doikou, A. Kundu, Nucl. Phys. B902 (2016), 415.

\bibitem{AvanCaudrelier}
J. Avan and V. Caudrelier, J. Geom. Phys. 120 (2017), 10.

\bibitem{Bazh}
V.V. Bazhanov and  S.M. Sergeev, J. Phys. A39 (2006) 3295;\\
V.V. Bazhanov, V.V. Mangazeev and  S.M. Sergeev, J. Stat. Mech.0807 (2008) P07004.

\bibitem{Bazh2}
V.V. Bazhanov and S.M. Sergeev, Nucl. Phys. B926 (2018) 509.

\bibitem{CauKu}
V. Caudrelier and A. Kundu, JHEP 02 (2015), 088  \\
V. Caudrelier and M. Stoppato, J. Geom. Phys. 148 (2020) 103546

\bibitem{CJM}
P.A. Clarkson, N. Joshi and M. Mazzocco, 
Théories Asymptotiques et Equations de Painlevé. Séminaires et Congrès (14). 
Sociètè Mathèematique de France, Paris, France, (2006) 53.

\bibitem{Degalomba1}
A. Degasperis and S. L. Lombardo, J. Phys. A 40 (2007) 961;\\
A. Degasperis and S. L. Lombardo, J. Phys. A 42 (2009) 385206.

\bibitem{DoikouFindlay1}
A. Doikou and I Findlay, PoS(CORFU2019)210.

\bibitem{DFS1}
A. Doikou, I. Findlay  and S. Sklaveniti, Nucl. Phys. B941 (2019) 361;\\
A. Doikou, I. Findlay and S. Sklaveniti,  Nucl. Phys. B941 (2019) 376. 

\bibitem{DoFioRa}
A. Doikou, D. Fioravanti and F. Ravanini, Nucl. Phys. B790 (2008) 465.

\bibitem{DoiSkl}
A. Doikou and S. Sklaveniti,  J. Phys. A 53 (2020) 255201.

\bibitem{DoikouXXZ}
A. Doikou, J. Stat. Mech., (2006) P09010.

\bibitem{DST} 
J.C. Eilbeck, P.S. Lomdahl and A.C. Scott, Physica D 16 (1985) 318;\\
A.C. Scott and J.C.  Eilbeck, Phys. Lett. A 119 (1986) 60.

\bibitem{FadTakRes}
L.D. Faddeev, N.Yu. Reshetikhin and L.A. Takhtajan, {\em Quantization of Lie groups and Lie
algebras}, Leningrad Math. J. 1 (1990) 193.

\bibitem{FT}
L. D. Faddeev and L. A. Takhtajan, {\em Hamiltonian Methods in the Theory of Solitons,}
(1987) Springer-Verlag.

\bibitem{Findlay}
I. Findlay, Physica D 398 (2019) 13.

\bibitem{Cluster}
 S. Fomin  and a. Zelevinsky, J.  Amer. Math. Soc. 15.  (2002) 497.

\bibitem{FordyKulish}
A.P. Fordy and P.P. Kulish, Commun. Math. Phys. 89 (1983) 427.

\bibitem{FredelMaillet}
 L. Freidel and J.M. Maillet, Phys. Lett. B263 (1991) 403.

\bibitem{GerIv}
V.S. Gerdjikov and M.I. Ivanov, Theor. Math. Phys. 52 (1982) 676

\bibitem{entr1}
J. Hietarinta, T. Mase and R. Willox, {\it Algebraic entropy computations for lattice
equations: why initial value problems do matter}, {\tt arXiv:1909.03232 [nlin.SI]}, (2019).

\bibitem{HJN}
J. Hietarinta, N. Joshi and F. Nijhoff, {\em Discrete Systems and Integrability}, 
Cambridge University Press, Cambridge (2016).

\bibitem{HV}
J. Hietarinta and C. Viallet, Nonlinearity 25 (2012) 1955.

\bibitem{Hirota}
R. Hirota, J. Phys. Soc. Japan 43 (1977) 1424.

\bibitem{HoneKouloukas} 
A.N.W. Hone, Ph. Lampe and Th. E. Kouloukas, 
{\it Cluster algebras and discrete integrability}, 
{\tt arXiv:1903.08335  [math.CO]}.

\bibitem{HoKou}
A.N.W. Hone and Th. E. Kouloukas, J. Phys A53 (2020) 364002.

\bibitem{Jimbo} 
M. Jimbo,  Commun. Math. Phys. 102 (1986) 537. 

\bibitem{KakMug}
F. Kako and  N. Mugibayashi, Prog. Theor. Phys. 61 (1979) 776.

\bibitem{Kedem}
R. Kedem,  J. Phys. A41 (2008) 194011.

\bibitem{Kono}
B.G. Konopelchenko, Phys. Lett. A87  (1982) 445.

\bibitem{Korepin}
V.E. Korepin, N.M, Bogoliubov and A.G. Izergin, 
{\it Quantum Inverse Scattering Method and Correlation Functions}, 
Cambridge University Press, Cambridge (1993)

\bibitem{Korff}
C. Korff,  J. Phys. A49 (2016) 104001.

\bibitem{Kulish} 
P.P. Kulish, Lett. Math. Phys., 5(3) (1981) 191.

\bibitem{KunduRagnisco} 
A. Kundu and O. Ragnisco, J. Phys. A27 (1994) 6335.

\bibitem{Sklyanin}
 V. B. Kuznetsov, M. Salerno and E. K. Sklyanin,  J. Phys. A33 (2000) 171. 

\bibitem{Manakov}
S. V. Manakov, Sov. Phys. - JETP 38 (1974) 248.

\bibitem{Darboux}
V. B. Matveev and M.A. Salle, 
{\em  Darboux transformations and solitons}, (1991) Springer-Verlag.

\bibitem{Mik1}
A.V. Mikhailov, Physica D3 (1981) 73.

\bibitem{NRGO}  
F.W. Nijhoff, A. Ramani, B. Grammaticos and Y. Ohta,  Stud. Appl.
Math. 106 (2001) 261.

\bibitem{Pap}
V.G. Papageorgiou, Yu.B. Suris, A.G. Tongas and A.P. Veselov, SIGMA 6 (2010) 033.

\bibitem{Rourke}
D.E. Rourke, J. Phys, A37 (2004) 2693.

\bibitem{Schiff}
J. Schiff, Nonlinearity 16 (2003) 257.

\bibitem{STS} 
M.A. Semenov-Tian-Shansky, Funct. Anal. Appl. 17 (1983) 259.

\bibitem{Sklyanincl}
E. Sklyanin, Funct. Anal. Appl.16(1982) 263;\\
E. Sklyanin, Preprint LOMI E-3-97, Leningrad, (1979);\\
E. Sklyanin; Zap. Nauch. Seminarov LOMI95 (1980), 55;

\bibitem{Sklyanin2}
E. Sklyanin, Funct. Anal. Appl. 21 (1987) 164;\\
E. Sklyanin, J. Phys. A21 (1988), 2375.



\bibitem{Ves}
A.P. Veselov, Phys. Lett.  A314 (2003) 214.

\bibitem{entr2}
C.M. Viallet, {\it Algebraic entropy for lattice equations}, 
{\tt math-ph/0609043}, (2006).

\bibitem{Yang} 
C.N. Yang, Phys. Rev. Lett.19 (1967) 1312.

\bibitem{Zab}
A. Zabrodin, Theor. Math. Phys. 113 (1997) 1347;\\
A. Zabrodin, Int. Journ. Mod. Phys. B11 (1997) 3125.

\bibitem{ZakharovShabat1}
V. E. Zakharov and A. B. Shabat, Sov. Phys. - JETP 34 (1972) 62.

\bibitem{ZakharovShabat2}
V.E. Zakharov and A.B. Shabat, Funct. Anal. Appl., 8, (1974) 226;\\
V. E. Zakharov and A. B. Shabat, Funct. Anal. Appl. 13 (1979) 166.

\bibitem{Zamo}
A.B. Zamolodchikov, Commun. Math. Phys. 79 (1981) 489.

\bibitem{Zullo}
F. Zullo, J. Math. Phys. 54 (2013) 053515.

\end{thebibliography}
\end{document}